\documentclass[usenatbib]{mn2e}
\usepackage{epsfig}
\usepackage{amsmath}
\usepackage{amssymb}
\title[Disks and Bulges of Group Galaxies]{Evolution
  in the Disks and Bulges of Group Galaxies since $z=0.4$}

\author[McGee et al.]{Sean L. McGee$^{1}$, Michael L. Balogh$^{1}$, Robert D. E. Henderson$^{1}$, David J. Wilman$^{2}$,
  \newauthor Richard  G. Bower$^{3}$, John~S. Mulchaey$^{4}$, Augustus Oemler Jr.$^{4}$
\\
$^{1}$Department of Physics and Astronomy, University of Waterloo, Waterloo, Ontario, N2L 3G1, Canada\\
$^{2}$Max--Planck--Institut f{\" u}r extraterrestrische Physik, Giessenbachstrasse 85748 Garching Germany\\
$^{3}$Department of Physics, University of Durham, Durham, UK, DH1 3LE\\
$^{4}$Observatories of the Carnegie Institution of Washington, 813 Santa Barbara Street, Pasadena, California, USA\\
}

\date{\today}
\def\ho{$h_{75}^{-1}$}

\begin{document}
\maketitle

\begin{abstract}

We present quantitative morphology measurements of a
sample of optically selected group galaxies at 0.3 $<$ z $<$
0.55 using the {\it Hubble Space Telescope  (HST)} Advanced Camera for Surveys
(ACS) and the \textsc{gim2d} surface brightness--fitting software
package. The group sample is derived from the Canadian Network for
Observational Cosmology Field Redshift survey (CNOC2) and follow-up Magellan
spectroscopy. We compare these measurements to a similarly selected
group sample from the Millennium Galaxy Catalogue (MGC) at 0.05 $<$ z
$<$ 0.12. We find that, at both epochs,  the group and field fractional bulge luminosity (B/T)
distributions differ significantly, with the dominant difference being a
deficit of disk--dominated (B/T $<$ 0.2) galaxies in the group
samples. 
At fixed luminosity, z=0.4 groups have $\sim$ 5.5
  $\pm$ 2 $\%$ fewer disk--dominated galaxies than the field, while by
  z=0.1 this difference has increased to 
  $\sim$ 19 $\pm$ 6 $\%$. Despite the morphological evolution we see no evidence that the
group environment is actively perturbing or otherwise affecting the
entire existing disk population. At both redshifts, the disks of group
galaxies have similar scaling relations and show similar median
asymmetries as the disks of field galaxies.  We do find evidence that the fraction
of highly asymmetric, bulge--dominated
galaxies is 6 $\pm$ 3 $\%$ higher in groups than
in the field, suggesting there may be enhanced merging in group
environments. We replicate our group samples at $z=0.4$ and $z=0$ using the semi-analytic galaxy
catalogues of \citet{bowermodel}. This model accurately
reproduces the B/T distributions of the group and field at z=0.1.
However, the model does not reproduce our finding that the deficit of
disks in groups has increased significantly since z=0.4.

\end{abstract}


\section{Introduction}
Galaxies, at a simple level, are a mixture of two fundamental and
distinct components: a bulge and a disk. In the local universe, bulge
dominated galaxies are generally red and quiescent, while disk
dominated galaxies are generally blue and actively forming stars
\citep{Blanton03}. Thus, the morphology of galaxies may be important when trying to
explain the observations that show the cosmic star formation density has
rapidly declined from a peak at z$\sim$1-1.5 \citep{lilly96, madau96,
hopkins2004}, and that the fraction of red galaxies has rapidly
increased over the same time \citep{faber, bell}. Indeed, these observations suggest that the process
which transforms galaxies from disk-dominated to bulge-dominated is
also the process which transforms them from the blue cloud to the red sequence
\citep[eg.,][]{Bell07}. However, studies of red, disk-dominated
galaxies \citep{wolf} and blue, bulge-dominated galaxies
\citep{91abraham, menanteau} suggest this model may be
too general. In addition, the large fractions of passive
spirals at intermediate redshift suggests that the truncation of
star formation may happen before the morphological transformation
mechanism (\citealt{pog_smail}, but see \citealt{wilmanMIPS}).

The local environment of a galaxy is an important factor in its
evolution. Observations of galaxy clusters have shown that galaxies
within clusters have lower star formation rates than
the general field \citep[eg.][]{balogh}. The rapid structure growth
associated with $\Lambda$CDM cosmology,
which increases the local galaxy density of the average galaxy with
time, may be the key driver of the decline of galaxy
star formation rates. Indeed, analogous to the lower star formation
rates in clusters, there is a correlation
between the local galaxy surface density and the morphology of
galaxies \citep{Dressler80}. At low redshift, the percentage of
early type galaxies increases, and the percentage of late types
decreases, with increasing density. Interestingly, \citet{Dressler80} found that this
relation was equally strong in centrally-concentrated, relaxed
clusters and in irregular, less centrally-concentrated clusters. At higher
redshift, z $\sim$ 0.5, \citet{Dressler97} showed that this  {\it
  morphology-density relation} is stronger in highly concentrated
clusters than in less concentrated clusters. 

Although these studies point to the crucial role clusters play in the
morphological transformation of galaxies, they are rare environments,
and thus cannot have a large enough effect on the properties of
galaxies to explain the decline of the star formation
density of the universe as a whole. However, the less dense
environment of optically selected groups is the most common environment
for galaxies in the local universe
\citep{2PIGG-cat}. Indeed, \citet{Postman} found that the
morphology-density relation extends smoothly into the group scale
environments. Further supporting the integral role of groups,
suppressed galaxy star formation rates in group-scale environments of
the local universe is now well-established
\citep[eg.,][]{balogh_ecology}. \citet{Wilman2} have shown that the
fraction of galaxies with [OII]$\lambda 3727$\AA\ emission, a measure of star formation, is much higher in group galaxies at intermediate
redshift, z $\sim$ 0.4, than in the local universe; however, the group
galaxies still exhibit suppressed star formation relative to the field
at the same epoch.   

The physical cause of the suppression of star formation since z$\sim$1 isn't
clear, but there are many candidates, each with their own morphological signatures. 
Within the context of $\Lambda$CDM cosmology, galaxy
mergers are often thought to be a dominant mechanism \citep{Hopkins07}. Simulations
suggest that a major merger between two gas-rich and star-forming
spiral galaxies produces a gas-poor, passively evolving elliptical
galaxy \citep{ToomreToomre, MihosHern}. If dominant, this scenario suggests that quiescent spiral galaxies should
be rare, and that the transformation of morphological type should
precede or happen at the same time as the complete suppression of star-formation. Group environments
are thought to be the ideal place for galaxy mergers because of their
high density and small relative velocities. 

Recently, driven by dual observations of large bubbles seen
in the hot X-ray gas of the intracluster medium \citep{McnamaraHydraA,
 Fabian00} and the correlation
between the mass of the galactic bulge and the size of the central
supermassive black hole \citep{Magorrian97,Ferrarese00}, feedback from
active galactic nuclei has
become a popular explanation for the suppression of star formation
rates in massive galaxies. Semi-analytic galaxy formation models have successfully
introduced these mechanisms in a parametrised way \citep{bowermodel,
  croton}, but the details are
still uncertain. Such energy feedback mechanisms may not
directly alter the galaxy morphology, but reduced star formation may result
in significant fading of the disk component. 

Meaningful morphological measurements are necessary
to break the degeneracy of physical explanations of star formation truncation.
Visual classification of galaxies onto a Hubble (or similar) system
has proven to be very useful for the study of galaxy
evolution. However, the high resolution and uniform quality of
large galaxy surveys has given rise to automated morphology
systems which attempt to make more quantitative measurements than a
visual system will allow. Non-parametric morphology systems \citep[eg.][]{abraham03,lotz} are robust, but
are not easily linked to physical quantities such as bulge or disk
scale lengths.
For this reason, in this paper we use a popular code, \textsc{gim2d} \citep{Gim2d},  to fit parametric
models to the surface brightness profiles. Parametric systems suffer
because they fit an {\it a priori} model to the galaxy surface
brightness and, as such, are prone to giving non-physical
results in some cases. 
We therefore adopt the 
logical filtering system proposed by \citet{mgc-gim2d}, to help
mitigate some of these effects.  
 
In this paper, we examine the morphological properties of optically
selected samples of group galaxies at z=0.4 and z=0.1. In
\textsection \ref{data} we describe our data samples and our
morphological measurements.  In \textsection \ref{result} we present the
main data results and in \textsection \ref{discuss} we discuss what
the data tells us about galaxies in transformation and compare our
data results with the semi-analytic galaxy catalogue of
\citet{bowermodel}.   We summarize our main results in \textsection
\ref{summary}. Details of the group-finding algorithm are given in
Appendix A, while in Appendix B we consider possible systematic effects
on our results. Finally, in Appendix C we show a representative sample
of images from our CNOC2 z=0.4 sample of galaxies. Throughout this
paper we assume a cosmology with matter density $\Omega_m$ = 0.3,
energy density $\Omega_\Lambda$ = 0.7, and present-day Hubble constant
$H_0$= 100$h$ km s$^{-1}$ Mpc$^{-1}$ with $h$ = 0.75 (or $h_{75}=1$).

\section{The Data} \label{data}

\subsection{The 0.3 $\leqq$ z $\leqq$ 0.55 Sample } \label{kcorrect}

Our moderate-redshift galaxy sample is derived from the Canadian Network
for Observational Cosmology Field Galaxy
Redshift Survey (CNOC2), a spectroscopic and photometric survey completed with
the  Multi-Object Spectrograph instrument at the 3.6-m
Canada France Hawaii telescope \citep{CNOC2}. The survey was
designed to study galaxy clustering and evolution. It targeted
galaxies in the redshift range 0.1 $<$ z $<$ 0.6 over four different
patches of sky totaling about 1.5 square degrees. The survey consists
of 5 colour (U,B,V,R$_C$, I$_C$) photometry of $\sim$40,000 galaxies to a
limiting magnitude of R$_C$=23.0 mag. Spectroscopic redshifts of
$\sim$6000
galaxies were obtained with an overall sampling rate of 48$\%$ to
R$_C$=21.5. This large survey allowed \citet{CNOC2-groups} to identify
a set of 200 groups using a friends-of-friends redshift-space group
finder. 

\citet{Wilman1} followed the CNOC2 survey with deeper spectroscopy of
a set of 26 groups (20 targeted, 6 serendipitous) drawn from the \citeauthor{CNOC2-groups} catalogue using the
Low Dispersion Survey Spectrograph (LDSS2)
at the 6.5m Baade telescope at Las Campanas Observatory  in
Chile. These groups were chosen to lie within 0.3$<$z$<$0.55, and
galaxies brighter than R$_C$=22.0 were targeted for spectroscopy. This additional spectroscopy was designed to
give near full completeness at bright magnitudes \citep[R$_C$$<$20; for details, see][]{Wilman1}. 

For each of the 20 targeted groups, we obtained single orbit {\it Hubble
Space Telescope (HST)} Advanced Camera for Surveys (ACS)
pointings in the F775W filter during Cycle 12. These
data were processed with the ACS pipeline as described by
\citet{ACS-pipe}. The images were further processed with the
$\it{Multidrizzle}$ task in $\it{pyraf}$ to remove cosmic rays and hot
pixels. 

Sources were detected in the {\it HST} ACS images using the SExtractor software
v2.3.2 \citep{sextractor}. For a source to be accepted, the signal in
at least 10 of its ACS pixels (0.5 arcsec$^2$) had to be a minimum of
1.3 $\sigma$ above the background. The faintest sources that are
reliably detected using these criteria have  R$_{775}$ $\approx$
23.9.  In this paper we restrict the analysis to sources with
redshifts (R$_C$ $<$ 22) and are therefore insensitive to these
detection parameters. However, we are sensitive to the deblending
parameters as the automated surface brightness fits use the
segmentation image produced by SExtractor to identify which pixels
belong to the galaxy. We used 32 deblending subthresholds, with a minimum
contrast parameter of 9.0 $\times$ 10$^{-4}$. By trial and error, these
parameters gave the best deblending upon visual inspection of the
output segmentation images.

\begin{figure}
\leavevmode \epsfysize=8cm \epsfbox{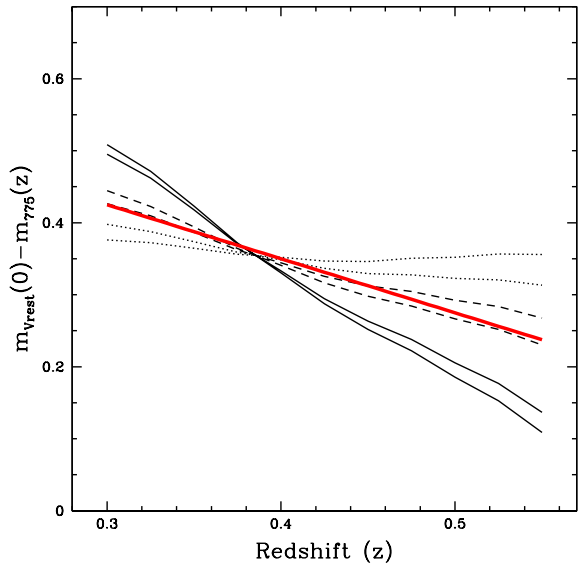}
\caption{We show k-corrections to rest frame V band magnitudes, derived from a suite of
  \citet{bc03} population synthesis models. The solid lines indicate a
  constant star formation rate; the dashed lines are models with an exponentially
  declining star formation rate; and the
  dotted lines represent single stellar population models created at $z=4$.  All models
  are shown with and without one magnitude of dust
  extinction.  The thick red line indicates the adopted k-correction of
   $k'=-0.75~z+0.65$.}
\label{kcor}
\end{figure}
We have used a suite of \citet{bc03} population synthesis models to
k-correct our {\it HST} magnitudes. We have chosen to avoid large
k-corrections by correcting all galaxies to the nearest restframe
waveband, $V$. 
Figure \ref{kcor} shows the models
used. The k-correction is mainly sensitive to the star
formation history and dust attenuation of the galaxy, and is
insensitive to the initial mass function and the metallicity. We add a
correction of $k'=-0.75~z+0.65$, shown by the red dashed line, to each
of our galaxies. This was chosen to minimize bias for any single
galaxy type, but dominates the uncertainty in the magnitudes to $\pm$
0.1. In all cases, our statistical uncertainties dominate over this source of
systematic error. All {\it HST} magnitudes are quoted in the k-corrected rest frame $V$ band.

\subsubsection{The Group and Field Samples} \label{cnoc_sample}

The group catalogue of \citet{CNOC2-groups} originally identified
virialized galaxy groups in redshift space using an iterative
friends-of-friends algorithm on the CNOC2 redshift catalog. They found
over 200 groups, with an average of 3.8 confirmed members per
group.  To take advantage of the deeper and more complete
LDSS2 spectroscopy available to us, we redefine the CNOC2 groups following
\citet{Wilman1}. We discuss the group finding algorithm in more detail in
Appendix \ref{lowfinder}. Briefly, an iterative procedure is used, which
initially selects galaxies within two times the velocity dispersion
of the mean group redshift, and with a transverse distance from the
group centre 
within 1/5 of the dispersion distance. In each iteration, the
velocity dispersion is recomputed using the Gapper estimator
\citep{Gapper} and the centre is recomputed as the luminosity-weighted
geometric centre of the group. The iterations are continued
until a stable group membership is reached.

Using these group centers and velocity dispersions, we restrict the
membership of our
group sample to galaxies within two velocity
dispersions of the group redshift and within 500\ho kpc of the group
center in the transverse direction. To obtain a true sample of group-sized halos we further restrict our group sample to have
velocity dispersions $<700$ km/s within 500\ho kpc of the group center. 

Field samples of galaxies are commonly defined in one of two ways: as an
``isolated'' sample, in which galaxies within groups and clusters are
removed, or as a ``global'' sample, which includes all galaxies
regardless of their environment. In practice, the removal of all
group and cluster galaxies is not possible in our sample. Incomplete
redshift sampling and the observational uncertainties associated with
group membership would lead to an ``isolated'' field sample which
still contained some group galaxies. Therefore, we prefer to
define a
field sample that contains all galaxies, regardless of their group or
cluster membership. 
However, our follow-up spectroscopy focused on regions that have
groups identified in the CNOC2 survey, and our morphologies are
derived from small ACS images centered on our each of 20 targeted
groups. To avoid the bias toward groups that would otherwise be
present, we define our field sample to include only those
galaxies that are not in groups as identified by the Carlberg et al. algorithm. As discussed in \textsection \ref{mgc_group} and
Appendix \ref{lowfinder}, because of the incompleteness of the CNOC2
survey, and the strict group finding algorithm of Carlberg, this
leaves a field sample which is only slightly depleted in group
galaxies when compared to the universe as a whole at z=0.4. A similar
selection in the semi-analytic group
catalogues discussed later (\textsection \ref{modresult}) shows that the field sample
will have only $\sim$ 8$\%$ fewer group galaxies than a true global field
sample. We could correct for this bias by creating a true
field sample which is an admixture made of 92$\%$ observed field sample and
8$\%$ group sample.  However, all of our conclusions are insensitive to this
correction, and for the sake of simplicity, we do not apply it to our results.  

Using these definitions yields a sample of 114 group
galaxies and 128 field galaxies with 0.3 $\leq$ z $\leq$ 0.55 and $M_V<-19$.


\subsection{The z $\sim$ 0.1 Sample }

Our sample of low redshift galaxies is derived from the Millennium
Galaxy Catalog (MGC). The MGC is a 37.5 square degree B-band imaging
survey carried out using the Wide Field Camera on the Isaac Newton
Telescope \citep{MGC-phot}. The survey is a long, 35 arcmin wide strip,
fully
contained within both the Two Degree Field Galaxy Redshift Survey
(2dFGRS) and the Sloan Digital Sky Survey (SDSS). The photometric
catalogs are complete to B$<$24 mag, and the imaging is of sufficient
quality to allow for the decomposition of galaxies into a bulge and
disk component. 

MGCz, the redshift survey component of the MGC, was designed to obtain
AAT/2dF spectra of B$<$20 galaxies which were not covered by either
2dFGRS or SDSS \citep{MGC-spectra}. This gives a redshift completeness of 96$\%$ for B$<$20
galaxies.

\subsubsection{The Group and Field Samples}\label{mgc_group}

There exist many low redshift group catalogues derived from either the
SDSS or 2dF surveys. However, we wish to create a catalogue that can be
compared directly and fairly with our higher-redshift group sample.  
The latter was derived from a
two-step process --- the initial survey and the targeted group
follow-up --- applied to an incomplete redshift survey.  
This method is not the most direct or efficient way to find groups in
our lower-redshift sample, but it does accurately 
reproduce our higher-redshift selection and the possible
biases within. We discuss the method in detail in Appendix
\ref{lowfinder}. Briefly, we reduce the completeness of the low
redshift sample to match the sampling rate of the CNOC2 survey, and then reproduce the \citet{CNOC2-groups}
algorithm. To mimic our follow-up of \citet{CNOC2-groups}
groups, we then increase the completeness and recompute the group membership. 

 Using this method we have found
19 groups with velocity dispersions between 100 km/s and 700
km/s and which lie in the range 0.06 $<$ z $<$ 0.12. Using a volume-limited sample of $M_B<-18$, we have a sample containing 99 group
members and 3022 field galaxies. As discussed in \S~\ref{cnoc_sample}, a true field sample would contain all
group members and field galaxies.  However, to maintain consistency between
the CNOC2 and MGC samples we do not exclude the small number of group
galaxies from our field sample. In practice, due to the size of the
field sample ($\sim$ 3000), adding or removing the 99 group galaxies has
no effect on the bulk properties of the field.
 
The low percentage of galaxies in groups in our sample seems in
contradiction with other group catalogues based on local redshift
surveys. For example, the 2dF Percolation Inferred Galaxy Groups (2PIGG) catalogue
\citep{2PIGG-cat}, which is based on the 2dF survey, finds that $\sim$
55$\%$ of galaxies are in groups in the local universe. However, the
fraction of 2PIGG groups with more than two members and within 
velocity dispersion limits of 100 km/s to 700 km/s (which are our
criteria) is only 24\%.
The reduction of the sampling rate to match the CNOC2
completeness results in the non-detection of about half of these groups.
A further 30$\%$ of galaxies are not found in groups because the
Carlberg algorithm is not a strict
friends-of-friends procedure: there is an additional step to check
that candidate group galaxies are overdense with respect to the
background.  Accounting for these selection effects, we would only expect to find 9 $\%$ of galaxies satisfying our definition
of a group.  Finally, we also remove
groups which are not fully contained within the very narrow MGC strip,
i.e. the group centers are within 500\ho kpc of a survey edge. This step reduces the
volume from which groups are selected from by $\sim$ 30 $\%$. Our group
catalogue is therefore
incomplete relative to 2PIGG, but our sample is robust (see
\textsection \ref{halos}) and accurately reproduces the CNOC2
group-finding algorithm at higher redshift. 
The small number of group
galaxies confirms, as suggested in \textsection \ref{cnoc_sample}, that
our field sample is only slightly depleted in group galaxies when
compared to the Universe as whole. Indeed, our results are unchanged if
we include these group galaxies in our field population at this
redshift, but for consistency with the CNOC2 groups, we do not. 

\subsection{Comparison of Surveys} \label{comparesur}

The measurement of a galaxy's morphology can depend on a number of
factors besides its intrinsic morphology, such as the imaging wavelength, angular
resolution and surface brightness limit. Because of these systematic
differences, direct measurement of morphological evolution is
difficult. In this paper, we largely concentrate on a direct
comparison between group and field galaxies at fixed redshift,
which eliminates the effects of these systematic differences. Nonetheless,
the MGC and CNOC2 are quite well matched in the key areas, which allows us
to compare the differential group and field behavior between redshifts.

Specifically, the two surveys probe approximately the same restframe wavelength.
The MGC survey is a rest frame B Band (observed frame $\sim$ 440 nm) survey, while our ACS
images are in the rest frame V Band (observed frame $\sim$ 775
nm). Since disks tend to be bluer than bulges one might expect a lower
B/T when measured in the B-band; however the intrinsic morphological differences are only
significant at much wider wavelength separation \citep{taylor}.

\begin{figure}
\leavevmode \epsfysize=8cm \epsfbox{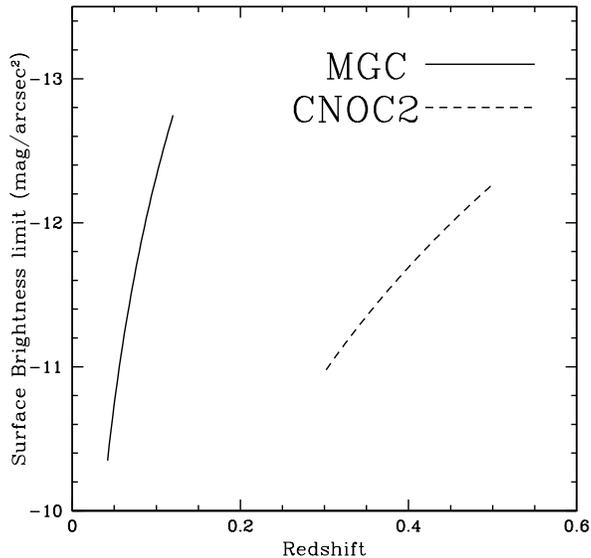}
\caption{A comparison of the absolute surface brightness limits of our
two samples of galaxies, as function of redshift. The solid line
represents the MGC surface brightness (26 mag/arcsec$^2$), while the
dashed line is the equivalent for the CNOC2 sample (30 mag/arcsec$^2$).}
\label{surf-bright}
\end{figure}

The apparent surface brightness limit of the MGC survey is 26
mag/arcsec$^2$ and that of the {\it HST} ACS images
is 30 mag/arcsec$^2$.  Figure \ref{surf-bright} shows the absolute
surface brightness limits of the two surveys as a function of
redshift, including (1+z)$^4$ cosmological dimming. For the redshift
ranges of interest, the absolute surface brightness limits are comparable. 

\begin{figure}
\leavevmode \epsfysize=8cm \epsfbox{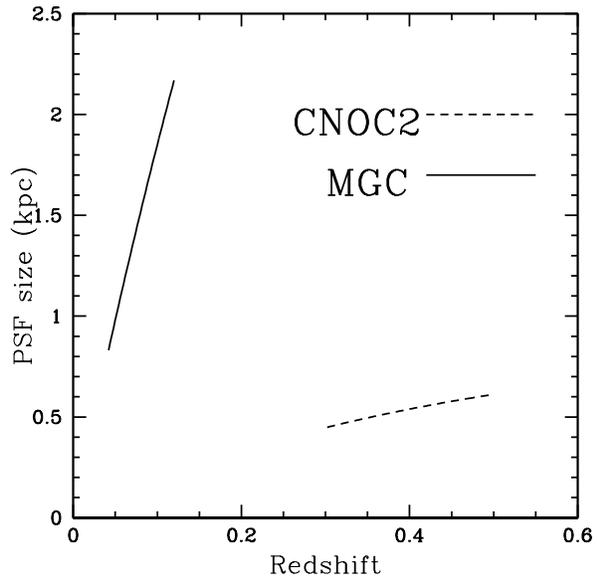}
\caption{A comparison of the point spread function FWHM of our
two samples of galaxies as a function of redshift. The solid line
represents the ground-based MGC PSF (1 arcsec), while
the dashed line corresponds to the {\it HST} resolution of the higher
redshift CNOC2 sample (0.1 arcsec).}
\label{psf}
\end{figure}

The excellent angular resolution of our {\it HST} ACS images (point spread
function (PSF) FWHM
$\sim$ 0.1 arcsec) allows for morphological measurements of the
CNOC2 sample. In fact, this gives a physical resolution which is
somewhat better than the MGC survey (PSF FWHM $\sim$
1 arcsec) in the redshift range of interest, as shown in Figure
\ref{psf}. Of greater concern is that the physical PSF of the MGC
survey has a large variation within the sample itself, due to the
redshift range spanned by the galaxies. In
Appendix \ref{mgc_red_depend}, we investigate the effect of the
physical resolution on the morphological measurements. We find that
the resolution differences have no effect on the
bulge-to-total light measurements, but do affect the asymmetry
parameter, causing the measured asymmetry to be higher in galaxies with better
resolution. In this paper, we only analyze asymmetry measures on
matched samples of galaxies, which mitigates this effect.

The properties of galaxies vary with luminosity \citep[eg.,][]{baldry}, so
when comparing morphological properties between the two surveys we
must use a common luminosity limit.  \citet{fukugita} have shown that late type galaxies (Scd to
Sab) have B-V=0.5--0.78, while early type galaxies (S0 and E) have
B-V=0.85--0.96. Thus, our $z=0.4$ magnitude limit of M$_V$$<$-19 corresponds
approximately to an MGC limit between $M_B=-18.5$ and $-18.0$,
depending on type and neglecting any luminosity evolution.  
Therefore, when directly comparing
the group and field behavior between the two surveys we only consider
MGC galaxies brighter than $M_B\approx -18$ (except in Figure
 \ref{diskfrac} which includes galaxies as faint as $M_B=-16$).

\subsection{\textsc{Gim2d} Morphological Measurements}\label{morph}

\citet{mgc-gim2d} have presented morphological measurements of the
MGC and we follow their procedure to derive morphological parameters
for the CNOC2 sample. We use the parametric IRAF package \textsc{gim2d} \citep{Gim2d,
MS}, which fits the sky-subtracted surface
brightness distribution of each galaxy with up to 12 parameters
describing a bulge and a disk component.  \textsc{gim2d} searches
the large-parameter space of models using a \citet{metro} algorithm,
which is inefficient but does not easily get trapped in local
minima. \citet{haussler} have shown that \textsc{gim2d} produces
  reliable fits with small systematic errors when the effective galaxy
  surface brightness is above the sky level, as it is for the galaxies
  in our sample. \footnote{Haussler et al also show that another parametric
  galaxy fitting code, \textsc{GALFIT} \citep{galfit}, performs better than
  \textsc{gim2d} in crowded fields. However, \textsc{GALFIT} uses a downhill gradient algorithm
which, although very efficient, may not be as robust as the simulated
annealing technique of \textsc{gim2d} for problems with many local minima, like two-
component fitting.}

The \textsc{gim2d} algorithm can fit single component (Sersic profile) or two component
(Sersic + exponential disk) models to the galaxy profile.
The Sersic profile is given by

\begin{equation}
I_b(R)=I_e \mathrm{exp}\left[-b_n[(R/R_e)^{1/n}-1]\right],
\end{equation}
where $I_e$ is the
intensity at the radius, $R_e$, and $n$ is the Sersic index. The
parameter $b_n$ is set to $1.9992n-0.3271$  within \textsc{gim2d} to
ensure that $R_e$ is the projected radius which encloses half the
total luminosity. 

In a two
component fit, the Sersic profile corresponds to the bulge model, and the disk is fit by an exponential model, 
\begin{equation}
I_d(R)=I_0 \mathrm{exp}(-R/h), 
\end{equation}
 where $I_0$ is the central intensity, $I_d(R)$ is the disk light
 profile as a function of radius $R$,and $h$ is the scale length.  

Ideally, we would like to fit two components to all galaxies, but often
galaxies do not have two resolvable components. It is for this reason
that we follow the prescription of \citet{mgc-gim2d}, who have made a
careful study of \textsc{gim2d} output. They suggest a logical filtering
system which initially fits a two-component model to the
galaxy. For galaxies which have ``normal'' light profiles,
this fit is kept, but for those galaxies which have
perturbed profiles or are obviously better described by a single component, we fit
a pure Sersic
profile. In practice, this has little effect on our results. 

\begin{figure*}
\begin{minipage}{0.45 \linewidth}
\leavevmode \epsfysize=8cm \epsfbox{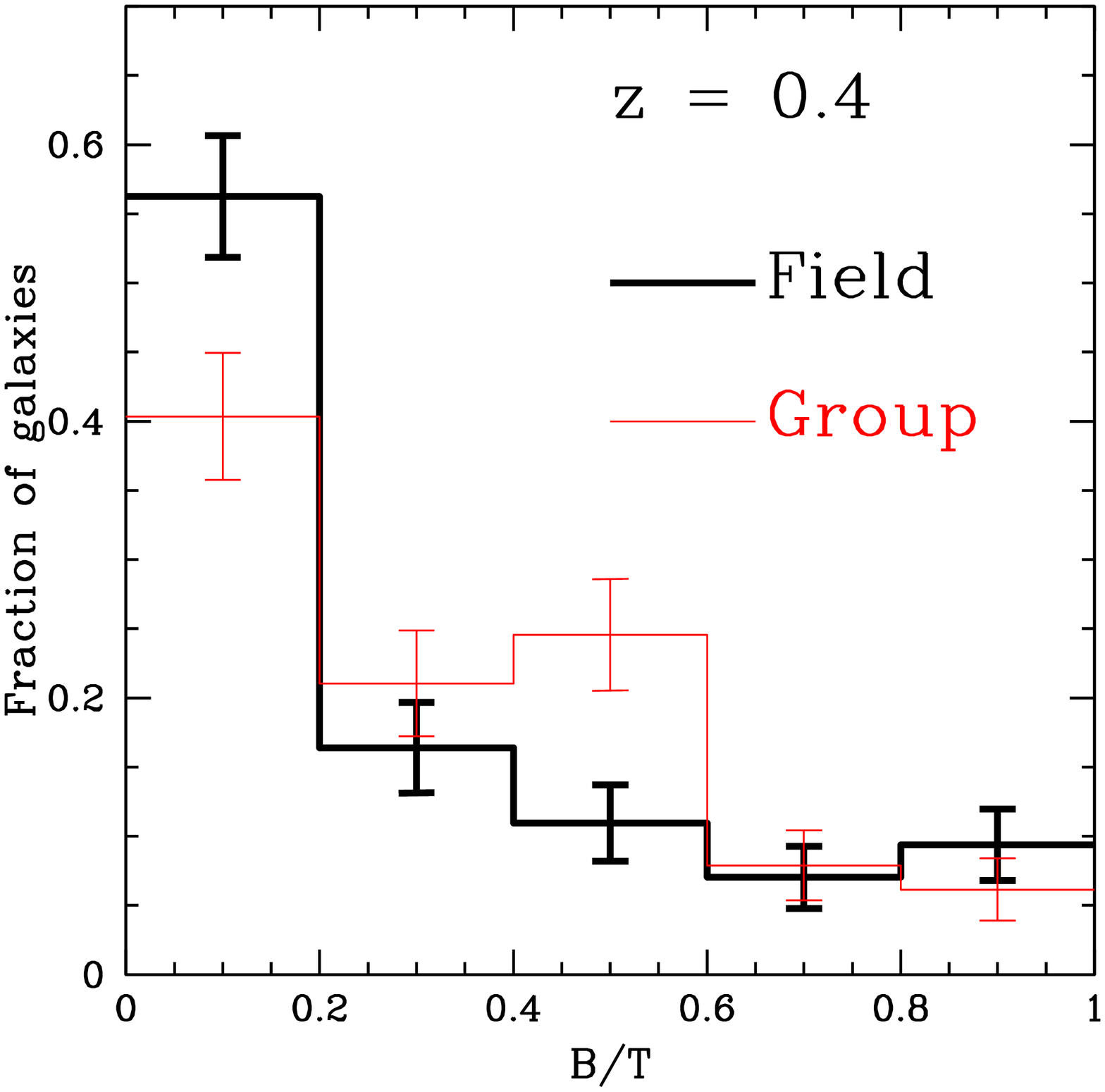}
\end{minipage}
\begin{minipage}{0.45 \linewidth}
\leavevmode \epsfysize=8cm \epsfbox{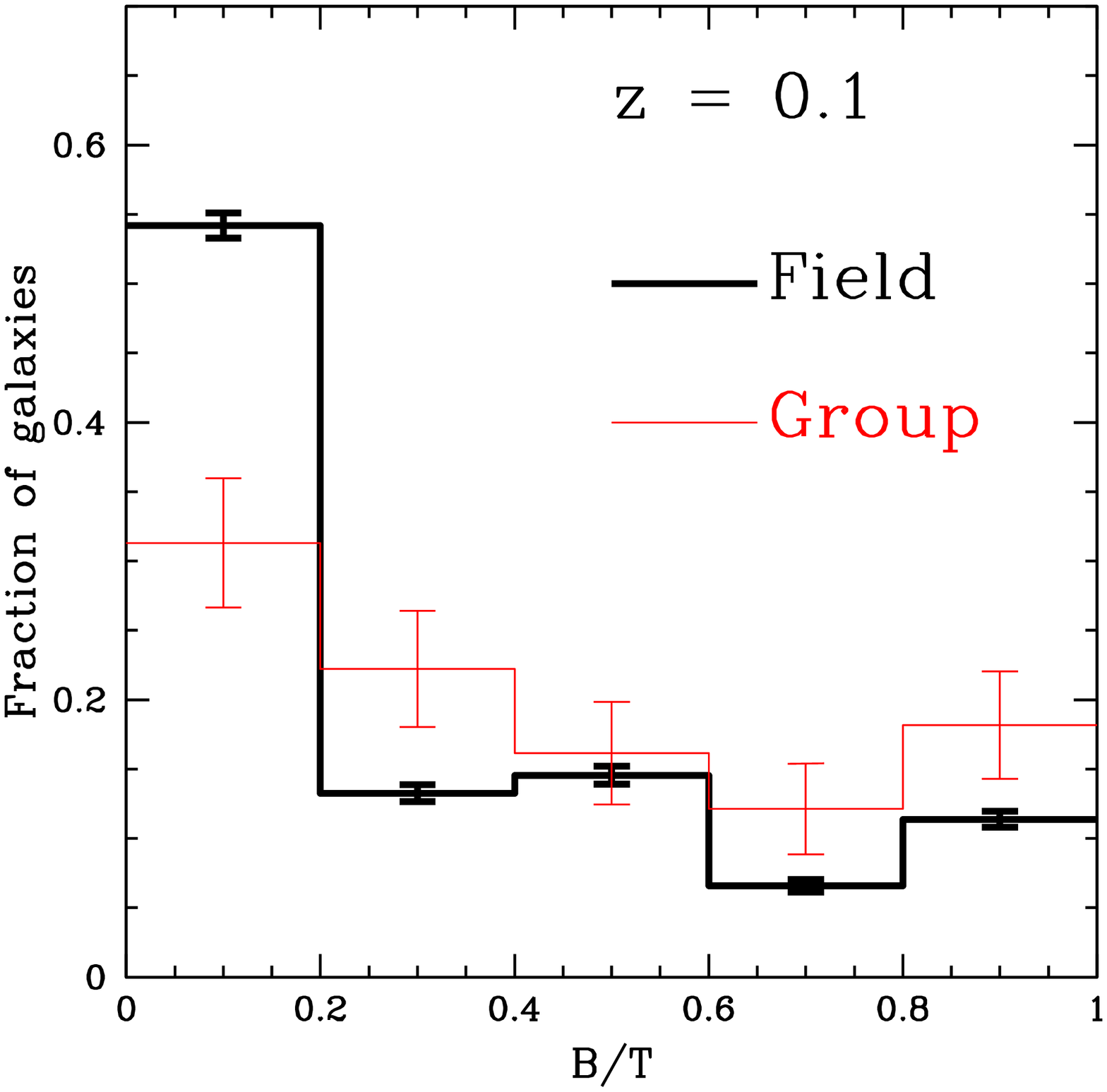}
\end{minipage}
\caption{The histogram of relative bulge luminosity for
the z $\sim$ 0.4 (left) and z $\sim$ 0.1 (right) samples. The thin red line
in each plot indicates the stacked group at that redshift, and the thick
black line is the field sample. Uncertainties are estimated with the
jackknife technique. }
\label{bt-cnoc}
\end{figure*}
\subsubsection{Residual Substructure} \label{Residual}
To quantify the substructure in the surface brightness profiles
we use the residual parameter, R, as defined by \citet{Schade1}. This parameter is also known as the
``asymmetry parameter, R'' \citep{Grothx} and the ``residual substructure
parameter, S'' \citep{McIntosh04}. It is defined as
\begin{equation}
R= R_T + R_A
\end{equation}
with 
\begin{equation}
R_T = \frac{\Sigma(|R_{ij} +R^{180}_{ij}|/2)}{\Sigma I_{ij}} - \frac{\Sigma(|B_{ij} +B^{180}_{ij}|/2)}{\Sigma I_{ij}}
\end{equation}
\begin{equation}
R_A = \frac{\Sigma(|R_{ij} -R^{180}_{ij}|/2)}{\Sigma I_{ij}} - \frac{\Sigma(|B_{ij} -B^{180}_{ij}|/2)}{\Sigma I_{ij}}
\end{equation}
where R$_T$ is the total residual parameter and R$_A$ is
the asymmetric residual parameter. R$_{ij}$ is the flux at pixel
position (i,j) in the residual image, R$^{180}_{ij}$ is the flux at
(i,j) in the residual image rotated by 180$^o$. B$_{ij}$ and
B$^{180}_{ij}$ are the corresponding values in the background
noise. Finally, I$_{ij}$ is defined as the flux at (i,j) in the object
image. The sum is done over all pixels out to r= 2r$_{hl}$, where
r$_{hl}$ is the radius at which half the galaxy's light is
enclosed.

\section{Results}\label{result}
Although our data sample is comprised of a relatively large number of group
galaxies (99 in the MGC sample and 114 in the CNOC2 sample), each
group has typically less than ten spectroscopically confirmed
members. Therefore, we stack the individual
groups to maximize the signal of the bulk group galaxy
proprieties. \citet{weinmann1} have shown that the velocity dispersion of a
group is a poor tracer of the mass of the group halo, especially for
groups containing few confirmed members. Although our group velocity
dispersions vary from 100 km/s to 700 km/s, \citet{Wilman1} has shown that the individual CNOC2 groups
show no significant differences based on group type or velocity dispersion
from a combined group. Thus we combine all the galaxies within 500\ho
kpc of a group center at each redshift to form a stacked z$\sim$0.1 group
and a stacked z$\sim$0.4 group. Because of the large uncertainties on the velocity
dispersions we do not attempt to estimate a ``virial radius'' for each
group, but instead simply require a group member to be within 500\ho kpc of
the group center, 
corresponding approximately to the expected virial radius of a typical
group in our sample, with a 360km/s velocity
dispersion.

\subsection{B/T Distribution}
In Figure \ref{bt-cnoc}, we present the quantitative morphology
distribution of the two samples. We compare field and group
distributions of the ratio of the bulge luminosity to total
luminosity (B/T) of each galaxy. Pure bulge galaxies have a B/T of 1,
while pure disk galaxies have a B/T of 0. The left hand panel of
Figure \ref{bt-cnoc} shows the B/T distribution for the z $\sim$ 0.4
sample of galaxies. The red line represents the
stacked group from the CNOC2 sample and the black line represents the field
galaxies.  A Kolmogorov-Smirnov (KS) test shows that the group and field
sample are not drawn from a common parent distribution, with 98.4$\%$ confidence.
The most significant difference within the CNOC2 sample is at B/T$<$0.2,
where the fraction of galaxies in the field ($\sim 57\pm 5$\%)
is higher than that in groups ($\sim 41 \pm 5$\%). In the right
hand panel of Figure \ref{bt-cnoc}, we present the B/T
distribution for the z $\sim$ 0.1 sample. Similar to the CNOC2 sample,
the fraction of B/T $<$ 0.2 galaxies is much higher in the field ($\sim
54 \pm 1$\%) than  in the groups
($\sim 32 \pm 6 $\%). A KS test rules out a common origin for the group
and field distributions of the MGC sample at greater than 99.9 $\%$ confidence. 

\begin{figure*}
\begin{minipage}{0.45 \linewidth}
\leavevmode \epsfysize=8cm 
\epsfbox{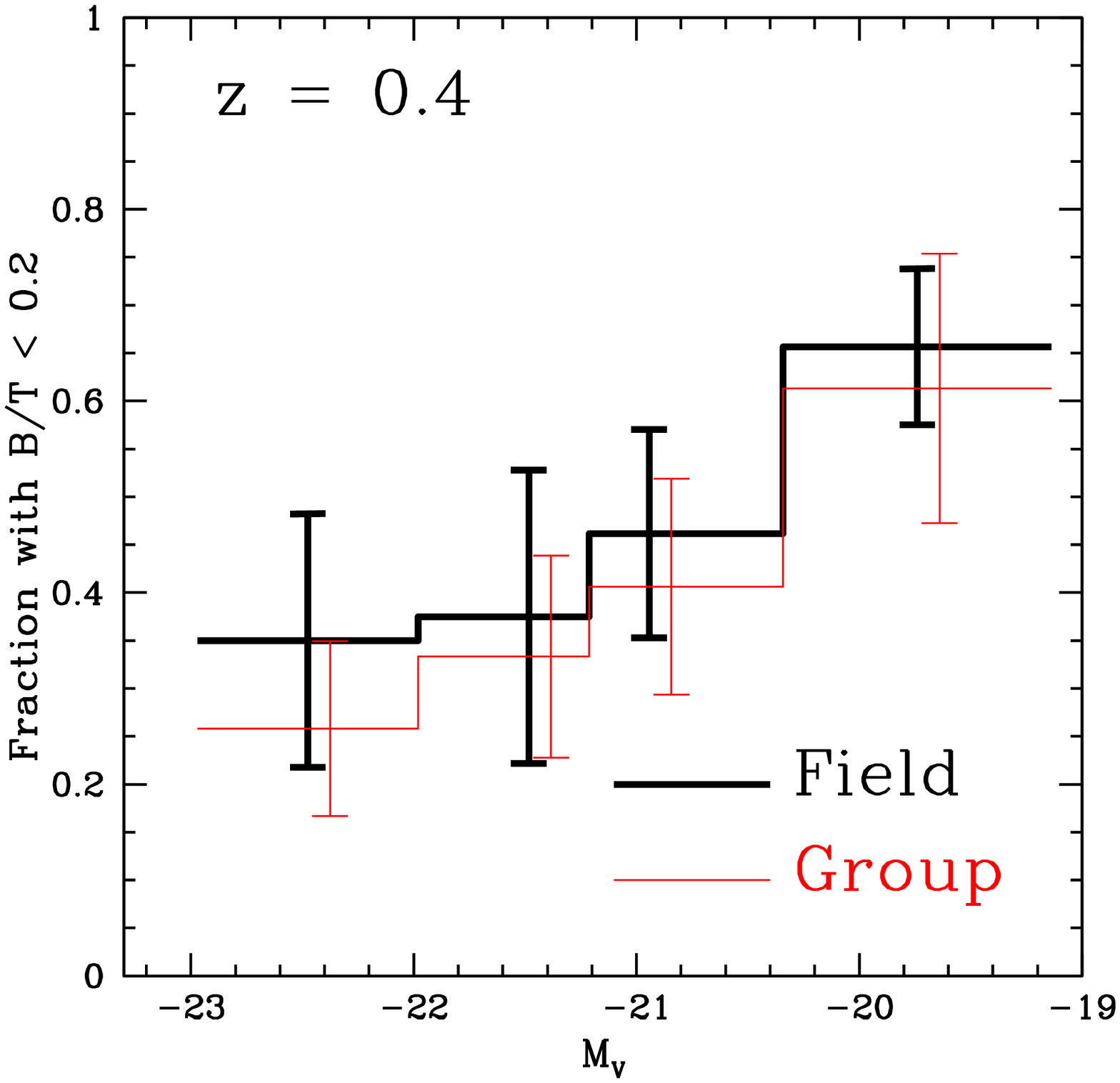}
\end{minipage}
\begin{minipage}{0.45 \linewidth}
\leavevmode \epsfysize=8cm \epsfbox{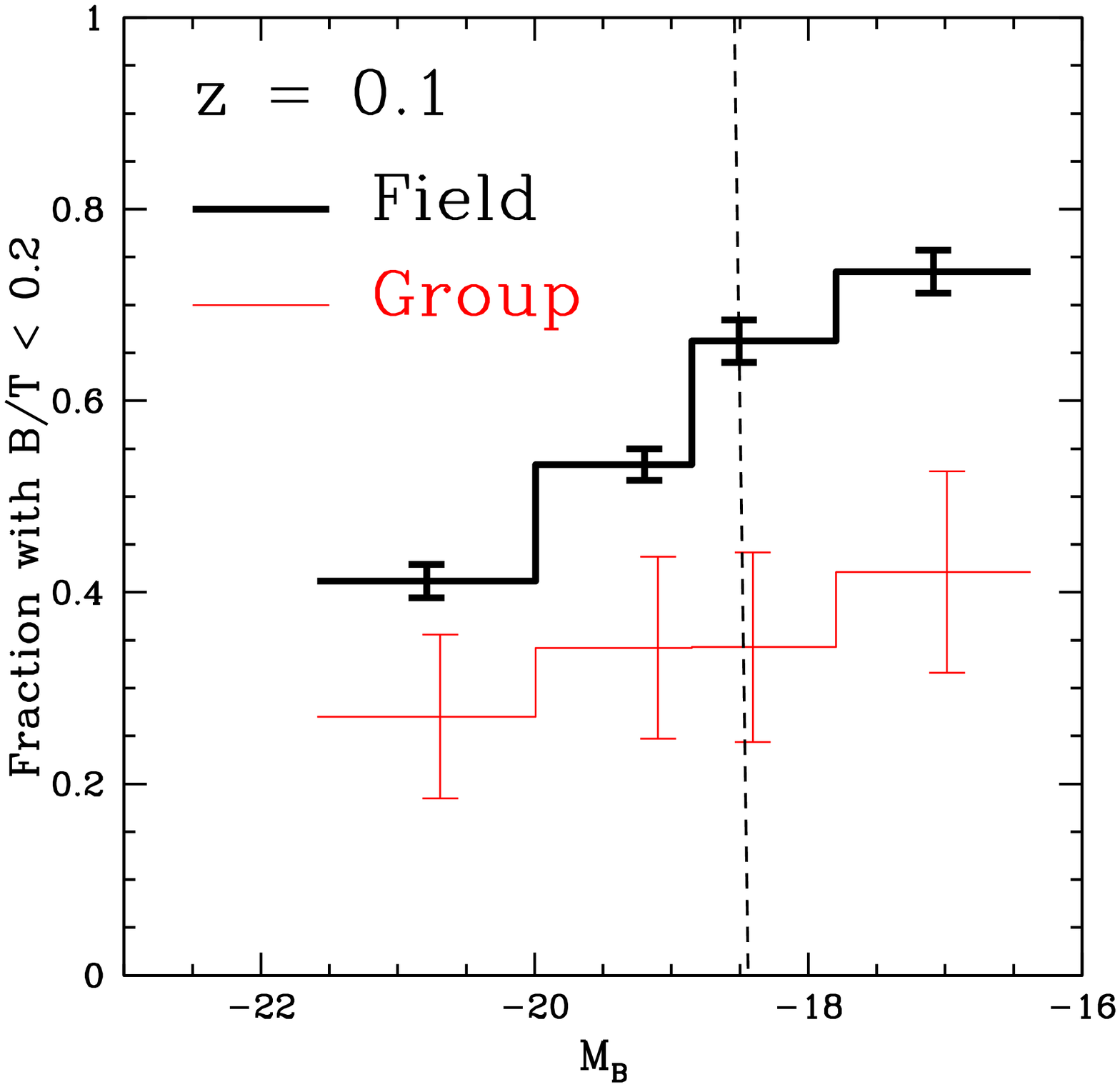}
\end{minipage}
\caption{The fraction of galaxies with B/T$<$0.2 (disk-dominated) as a
  function of absolute magnitude. Left: The z $\sim$ 0.4
  sample. Magnitudes are measured using the flux calculated by
  \textsc{gim2d} and k-corrected to rest frame V band. Right: The z $\sim$ 0.1
  sample, where magnitudes are measured in the B band. The field galaxies are represented by the
  thick black line and the group galaxies by the thin red
  line. The dashed vertical line indicates the equivalent B-band limit
  corresponding to the CNOC2 $M_V=-19$ magnitude limit, assuming a B-V color of
  0.63. Uncertainties are measured using the jackknife technique. }
\label{diskfrac}
\end{figure*}
In both plots we have seen evidence for the well-known
  morphology-density relation.  In this case, it is manifested as a deficit of disk dominated
  galaxies in the group samples when compared to the
field. However, it is known that bright galaxies tend to be more
frequently bulge-dominated than faint galaxies. Therefore, it
is possible that this form of the morphology-density relation is
related to different field and group luminosity distributions rather
than any intrinsic difference between group and field
galaxies at fixed luminosity. To explore this we present Figure
\ref{diskfrac}, which shows the fraction of disk-dominated galaxies as
a function of total galaxy magnitude. Studies of the B/T distribution
of visually classified galaxies have shown that elliptical
and S0 galaxies predominately have B/T$>$0.3-0.4 (\citealt{tran} and \citealt{wilmanMORPH}), but we chose
to define ``disk dominated'' to indicate galaxies with B/T$<$0.2. This
choice is made to isolate those galaxies in the bin with the most
significant difference between the group and field samples, but our conclusions are
unchanged even if we use B/T$<$0.4 to define disk-dominated galaxies. 

The left hand panel of Figure \ref{diskfrac} shows the z
$\sim$ 0.4 sample, with the thick black line representing field
galaxies and the thin red line representing the groups.  In any one
luminosity bin there is no
significant difference (ie.$>$ 1 $\sigma$) between the fraction of disk
galaxies in the group and field; however, overall there is a systematic
difference of 5.5 $\pm$ 2 $\%$, in the sense that the fraction of
disk-dominated galaxies in groups is always lower than in field
galaxies of comparable luminosity.

The right hand panel of Figure \ref{diskfrac} shows the disk fraction as a
function of magnitude for  the $z \sim$ 0.1 sample. The magnitude is the
rest frame B band, the waveband in which the B/T decomposition was
done. In this figure, we show all galaxies in our redshift
range to $M_B$=-16, fainter than our volume-limited sample of
$M_B$$<$-18. The dashed vertical line indicates the
equivalent B-band limit of the CNOC2 $M_V=-19$ magnitude limit, assuming a
B-V color of 0.63, which is typical of a late-type galaxy
\citep{fukugita}. At this redshift there is clearly a significant
difference in the fraction of disk-dominated systems between group and
field galaxies of the same luminosity. This difference is 24 $\pm$ 6 $\%$ over the
full magnitude range of the sample, and 19 $\pm$ 6 $\%$ brighter than the equivalent
CNOC2 luminosity limit.

Although the B/T distributions of the group and field galaxies are
different at both redshifts, this predominately reflects a difference in luminosity
distributions at high redshifts, but an intrinsic difference in the
fraction of disk galaxies of fixed luminosity at low redshift. It therefore appears that the morphological
segregation in groups has increased significantly from z $\sim$ 0.4 to
z $\sim$ 0.1.
We recall that the z=0.1
sample is measured in the rest frame B band while the z=0.4 sample is
measured in the 
rest frame V band.  It would be possible to mimic our results if the disks of group
galaxies were significantly redder than the disks of field
galaxies. However, to create a B-band disk fraction consistent with
the V band disk fraction, the disks of group galaxies must be B-V=2.07
redder than the disks of field galaxies. On average, in the MGC sample, pure
disk (B/T=0) group galaxies are only B-V=0.01 redder than the pure
disk field galaxies. Thus, it appears the evolution is real.  
We will further
explore the colours of these galaxies, including a comparison with the CNOC2 sample, in
a future paper. 

\subsection{Structural Parameters}\label{struct}
\begin{figure*}
\begin{minipage}{0.45 \linewidth}
\leavevmode \epsfysize=8cm \epsfbox{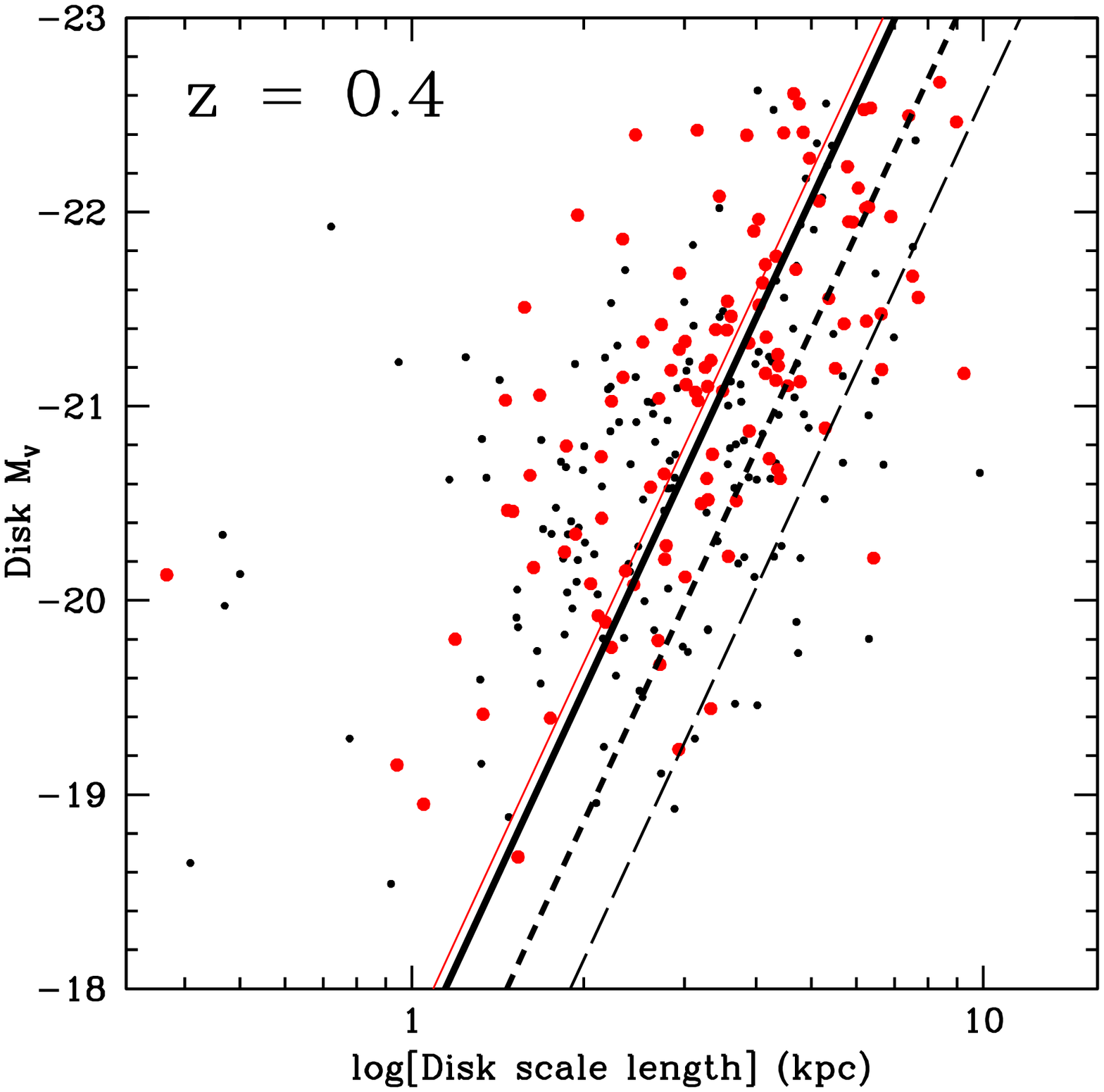}
\end{minipage}
\begin{minipage}{0.45 \linewidth}
\leavevmode \epsfysize=8cm \epsfbox{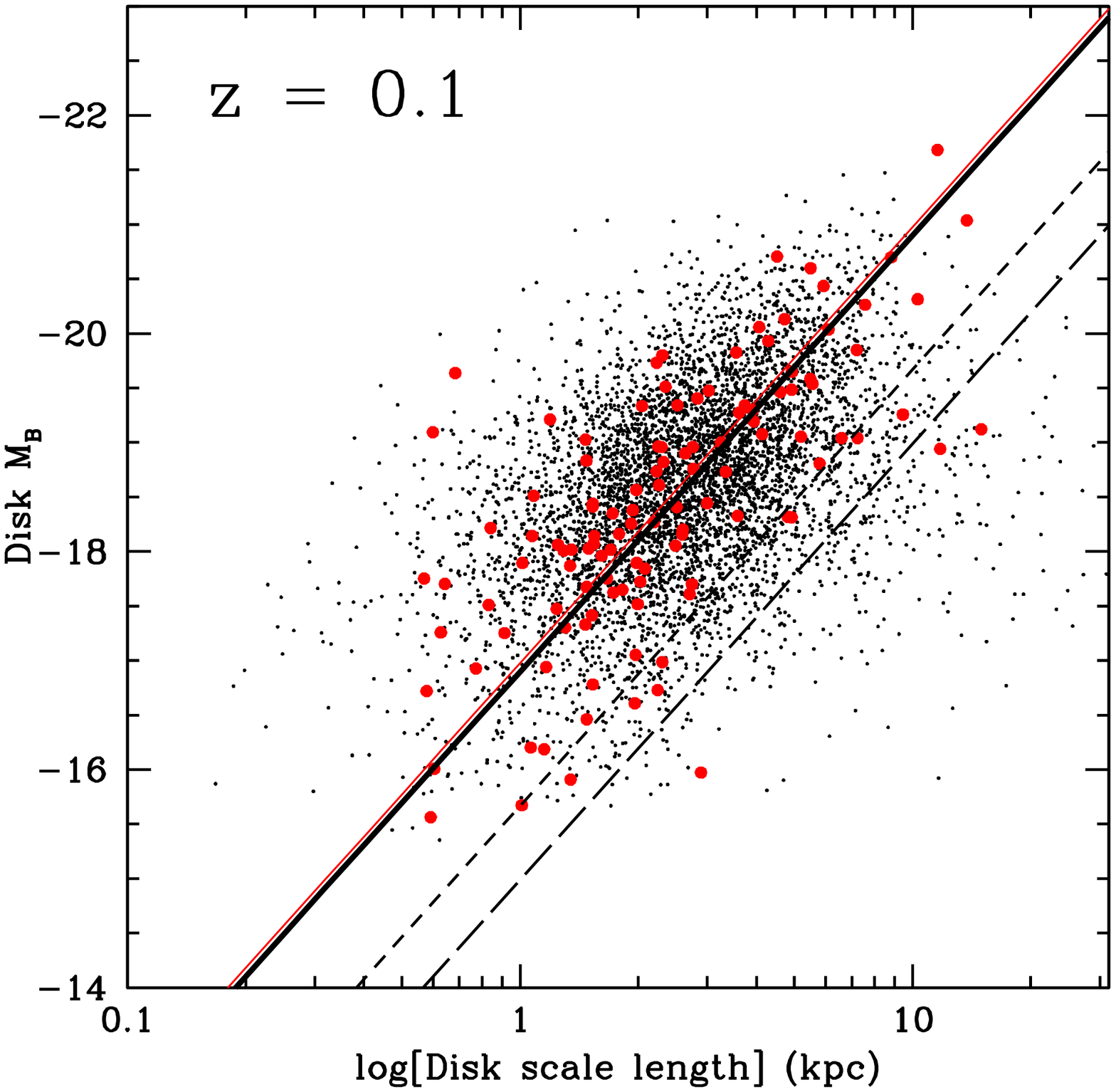}
\end{minipage}
\caption{The distribution of disk magnitudes as a function of disk
  scalelength for non-bulge dominated galaxies (B/T $<$0.7). Left: z $\sim$
  0.4 sample.  Right: z$\sim$ 0.1 sample. Large red points are group galaxies
  and small, black points are field galaxies. The thick solid black line is
  the best fit to the field, thin red line is the group best fit. The
  dotted (dashed) black line shows the effect of truncating star
  formation in field galaxies 1(3) Gyrs ago. }
\label{diskrad}
\end{figure*}
The evolution in morphological segregation suggests there is a change
occurring within the group galaxies between z $\sim$ 0.4 and z $\sim$
0.1. Here we focus on the possible causes of this
evolution. We first look
for structural differences between the group and field galaxies. If
star formation in group galaxies is quenched, then the disk of the
galaxy may slowly fade away, and we might expect to see a departure
from the normal scaling relations between disk size and luminosity.  

In Figure \ref{diskrad} we plot the distribution of disk magnitudes as
a function of the disk scalelength, excluding only the most bulge-dominated galaxies
(B/T $>$ 0.7). \footnote{An occasional problem in modeling the surface
brightness profile of galaxies with automated programs is the
tendency to fit small disk (or bulge) components to galaxies which may
have (e.g.) twisted isophotes. Thus we restrict our analysis to
galaxies which have significant disk (or bulge) components when
looking at their scaling properties.} The left hand panel shows the z
$\sim$ 0.4 sample, while the right hand
panel shows the z $\sim$ 0.1 sample. In both
plots it is evident that there is a correlation
between the disk size and its brightness, such that brighter disks
have larger disk scale lengths. This is not surprising, but the fact
that the group and field lie on the same relation (although with large scatter)
is. The thick solid black line is the best fit to the scaling relation of
the field galaxies. Best fit lines are determined using a robust
biweight estimator which minimizes the effect of distant outliers. In
principle, uniform fading of a perfectly exponential disk would result
in a lower disk luminosity, but would leave the scalelength
unchanged.\footnote{\citet{haussler} have shown that the recovered
scale length is underestimated for low S/N galaxies which are fit with a Sersic
profile. However, our data are sufficiently deep to avoid this problem, typically reaching
the surface brightness limiting isophotal radius at 3-5 disk scale lengths.
} Therefore, an ideal
population of faded disk galaxies would exhibit the same scaling relation, but with
different normalization. Adopting this assumption, we fit only the
normalization to the group galaxy relation, while maintaining the slope
defined by the field galaxies, as shown by the thin red line. There is
no statistically significant
difference between the normalizations of the group and field
populations (a difference of 0.086 $\pm$ 0.134
for the z=0.4 sample, and 0.103 $\pm$ 0.151 for the z=0.1 sample). 

\begin{figure*}
\begin{minipage}{0.45 \linewidth}
\leavevmode \epsfysize=8cm \epsfbox{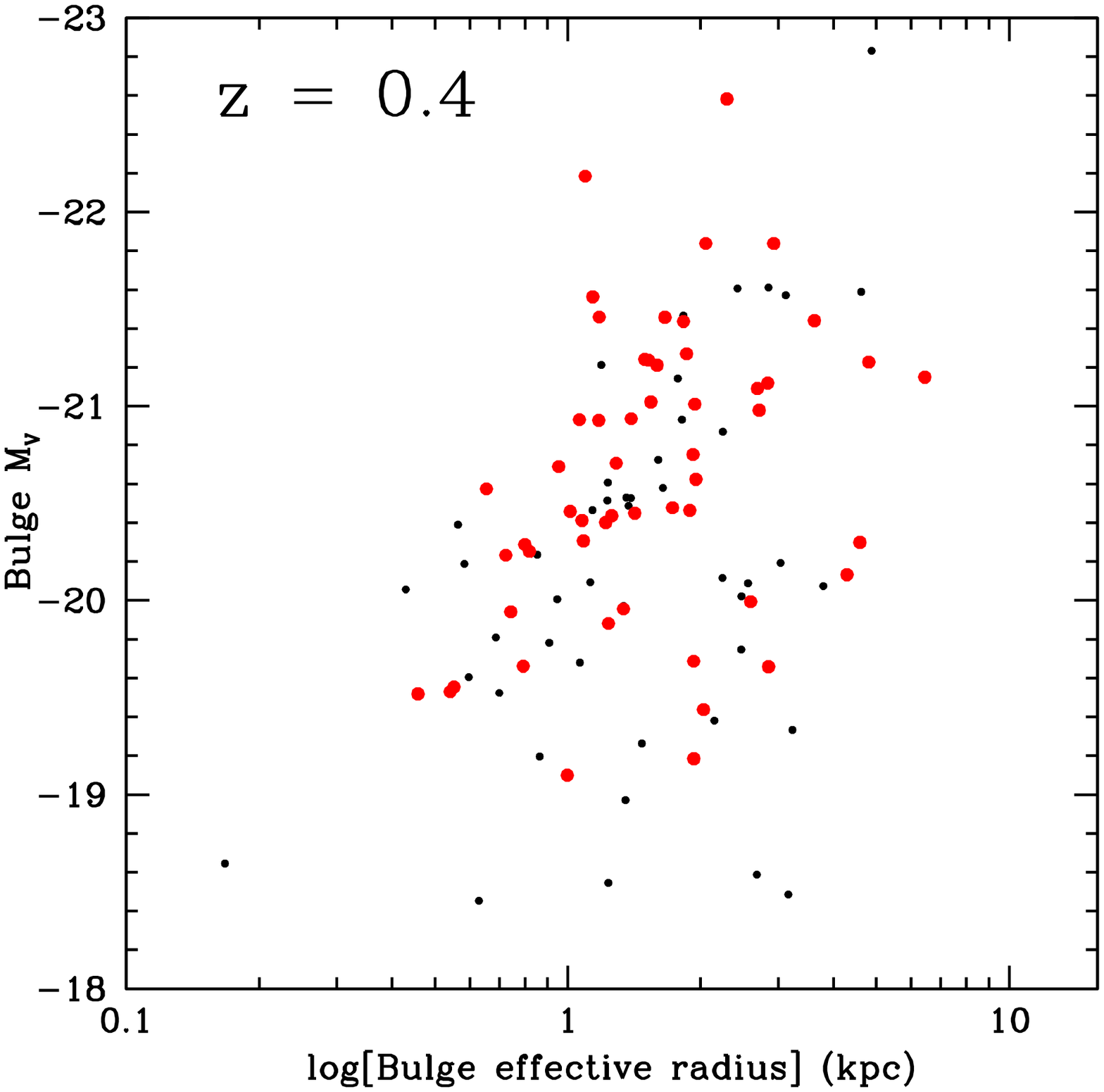}
\end{minipage}
\begin{minipage}{0.45 \linewidth}
\leavevmode \epsfysize=8cm \epsfbox{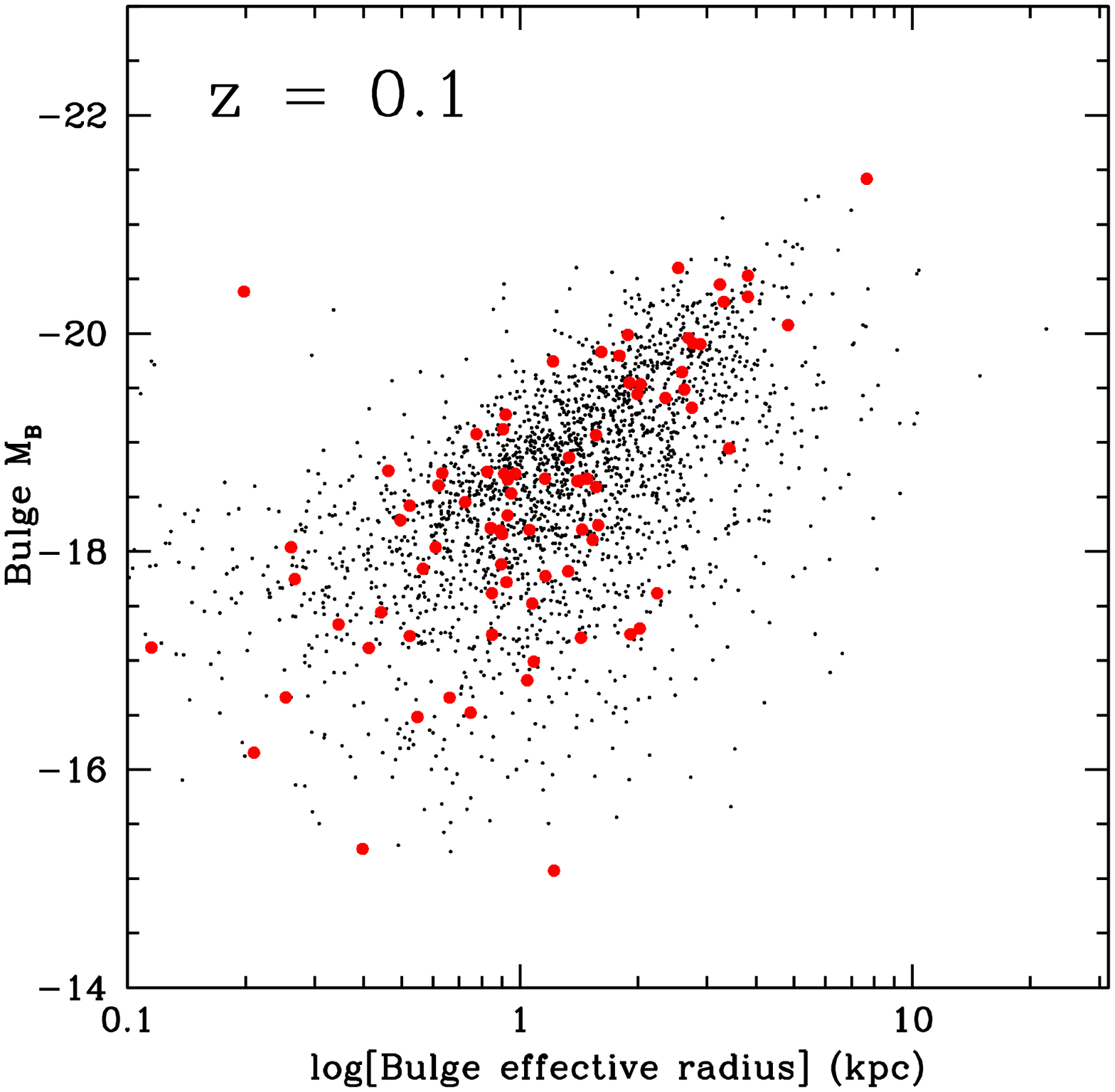}
\end{minipage}
\caption{The distribution of bulge magnitude as a function of bulge
  half light radius for non-disk dominated (B/T $>$ 0.3) galaxies. Left:
  The z$\sim$ 0.4 sample. Large, red points are group galaxies and small, black
  points are field galaxies.  Right: The z$\sim$ 0.1 sample. Red, filled points are group
  galaxies and black points are field galaxies.}
\label{bulgerad}
\end{figure*}

Since we do not see any significant difference in the disk scaling
relations for group galaxies, it
may be that the process of morphological transformation is instead dominated
by a growing bulge (for example, through mergers). Again, if this were true, we might expect to see a
deviation in the group and field bulge scaling relations. 
In Figure \ref{bulgerad}, we plot the distribution of bulge magnitudes
as a function of the bulge half light radius, excluding the most disk--dominated
galaxies (B/T $<$ 0.3). The left hand panel shows the z $\sim$ 0.4
sample and the right hand panel has the z$\sim$0.1 sample. The
field and group distributions are again similar, at both redshifts. In \textsection \ref{discuss}, we examine the constraints
these findings place on the amount of fading which is possible in the
group sample.

\subsection{Asymmetry} \label{asym}
In this section we examine the asymmetries of the galaxies, to
try to untangle a merger--driven transformation scenario from a gradual disk fading
model. 
Galaxy mergers and harassment often produce
noticeable asymmetric features, like tidal tails, 
which could manifest
themselves as deviations from the smooth
surface brightness profiles we have used for the morphological
measurements. If galaxy mergers were enhanced in groups, we
would expect to see an increase in the fraction of galaxies with large
asymmetries.  
On the other hand, the cessation of star formation in the disk would
likely result in the disappearance of bright clumps and spiral arms,
thereby removing residual substructure, making the disks appear
smoother. 

We have seen that the fraction of disk galaxies depends
on magnitude, and that the difference in the high redshift B/T
distribution is partly due to different luminosity distributions
in the field and group samples. For this reason, in the rest of this 
section, we match the field and group luminosity and redshift
distributions. In the MGC sample, we match each group galaxy to the
field galaxy with the closest magnitude and within 0.03 in
redshift. The magnitude differences for these matched galaxies are all less than 0.02
because of the large number of available field galaxies in the
sample. Similarly, for the CNOC2 sample, we match group galaxies to the
nearest field galaxy in magnitude (within 0.2 mags) and within 0.05
in redshift. We match all 99 group galaxies in the MGC and 105 of the
114 CNOC2 group galaxies. The brightest CNOC2 group galaxies have no field
counterpart, and so are effectively excluded from the remainder of this
analysis. 

\begin{figure}
\leavevmode \epsfysize=8cm \epsfbox{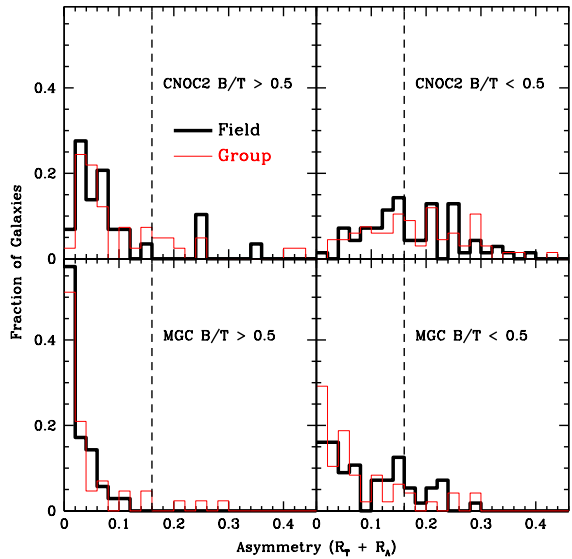}
\caption{The distribution of asymmetries (R$_T$+R$_A$) for CNOC2 bulge
  dominated galaxies (top left), CNOC2 disk dominated galaxies (top
  right), MGC bulge dominated galaxies (bottom left), and MGC disk
  dominated galaxies (bottom right). In all plots group galaxies are
  represented by the thin red line and the luminosity matched field is the thick black
  line. The dashed vertical line indicates the lower limit for
``highly asymmetric'' galaxies ($R_T~+~R_A >$ 0.16) as defined by \citet{pattonasym}.}
\label{asymetry}
\end{figure}

To probe the substructure of the galaxies, we have calculated the
asymmetry parameter according to the definitions given in \textsection
\ref{Residual}. In Figure \ref{asymetry} we show the distribution in
asymmetry for bulge dominated (B/T$>$0.5) and disk dominated
(B/T$<$0.5) galaxies for both the CNOC2 and
MGC surveys. The dashed vertical line indicates the lower limit for
``highly asymmetric'' galaxies ($R_T~+~R_A >$ 0.16), as defined by \citet{pattonasym}.
Disk
dominated galaxies have much higher median asymmetries than bulge
dominated galaxies, as expected.  However, there is no appreciable
difference in the median asymmetry between the 
group and luminosity-matched field, in any of the samples. 
Although we can not resolve the disapperance of individual HII regions, visual
  inspection of disk dominated galaxies in Figures \ref{group_thumb}
  and \ref{field_thumb} clearly show that galaxies with low B/T and
  high asymmetries have strong asymmetric structures typical of star forming
galaxies, ie. large spiral arms and lumpy regions. The lack
of a systematic difference in the asymmetries of matched group and field
disk-dominated galaxies means that we find no evidence for a mechanism
that suppresses star formation in group disks. 

We note that, as shown in Figure \ref{psf}, the CNOC2 sample has a
lower physical PSF than the MGC sample. We show in Appendix \ref{mgc_red_depend} that
galaxies which are blurred to have a lower physical resolution have
lower asymmetry values. Therefore, we do not consider the change in
median asymmetry between z $\sim$ 0.4 and z $\sim$ 0.1 as evidence of real
evolution.

\section{Discussion}\label{discuss}

\subsection{Galaxies in Transformation?}

In this section we investigate how our results in \textsection
\ref{result} might constrain the number galaxies that are in the process
of transforming within the group environment. Specifically, we have shown
in \textsection \ref{struct} that the bulge- and disk-components appear
to obey scaling relations between
size and luminosity that are independent of environment.  This
suggests that any transformation mechanism must either leave these
scaling laws intact, or only affect a small number of galaxies at the
epoch of observation.  To
investigate this, we consider a simple
\citet{bc03} model with a constant star formation rate as
appropriate for the disk components.  If
this star formation is suddenly truncated, within 1 Gyr it will have
subsequently faded by 1.23 magnitudes in rest-$B$ (as measured for the MGC
sample), or 0.68 magnitudes in rest-$V$ (appropriate for the higher-z
CNOC2 sample).  After 3 Gyr, the amount of fading expected is 1.91 mags (B) or
1.39 mags (V). However, such truncation should not affect the measured
scale length of the disk. Therefore, if star formation were truncated
in the entire population of group galaxies, we would expect the
normalization of the scaling relation to change by these amounts.
These relations are shown by the dotted
(1 Gyr) and dashed (3 Gyr) black lines in Figure
\ref{diskrad}. These lines are significantly offset from the measured group
relation (which is consistent with that of the field); therefore, we can
easily rule out that star formation has been recently truncated in the
entire disk population.

We would next like to establish what {\it fraction} of group galaxies could have
undergone 1 (3) Gyr fading and still have a scaling relation that is
consistent with the observed field scaling relation (ie. $<$ 2 $\sigma$
difference in the normalization, corresponding to a difference of 0.2 magnitudes
for the MGC sample, and 0.18 magnitudes for CNOC2). To assess this we
randomly choose a sample of group galaxies from Figure~\ref{diskrad} and fade their disks for
1 (3) Gyrs according to the Bruzual $\&$ Charlot model described
above. We recompute the
B/T ratios of the sample, and impose our selection criteria on total magnitude
and B/T (recall we are excluding the most bulge--dominated galaxies
from this analysis). 
 The normalization of the scaling relation is then refit to the
new set of data, keeping the slope fixed. We can then determine an
upper limit on the fraction of galaxies which may have undergone such
fading for 1 (3) Gyrs; these limits are 41 $\pm$ 3 $\%$
(29 $\pm$ 4 $\%$) for the CNOC2 sample, and 9 $\pm$ 3 $\%$ (4
$\pm$ 2 $\%$) for the MGC sample. The upper limit for the z=0.4 sample is quite high, as expected given
the relatively small sample size.  Therefore, based on these data alone, we cannot rule out the
hypothesis that a substantial fraction of these galaxies are undergoing
a significant change in their star formation rate.  The upper limit for the $z=0.1$
sample is much more restrictive, partly because of the sample size and
partly because the rest-frame B-band is more sensitive to recent star
formation.  Our limits mean that any mechanism able to truncate star
formation in disks is not dominant in present day groups.

Mergers are likely to be rare, but transformative events, so an
increased merger history in groups would not be expected to increase the
median asymmetry of group galaxies, but rather increase the fraction of highly
asymmetric galaxies. By examining the fraction of galaxies with
$R_T~+~R_A >$ 0.16, we see that an extra $\sim$ 6 $\pm$ 3 $\%$ of group galaxies are
highly asymmetric in the bulge-dominated samples at both redshifts,
compared with the 
matched field sample (\ref{asymetry}). The fraction of highly asymmetric galaxies in
the disk-dominated samples is similar for both the group and the luminosity-matched field.  
This may indicate that there is an increase of merging or
interacting galaxies in groups. A visual inspection of the CNOC2
galaxies shows that 4 out of 9 of the ``high-asymmetry'' bulge dominated
galaxies are indeed merging or interacting, as shown in Appendix \ref{thumbs}.

\subsection{A comparison with X-ray selected groups}
In \textsection \ref{asym}, we have shown that the median asymmetry of
group and field galaxies are statistically indistinguishable. These results are
interesting because \citet{tran} have shown that, in a
sample of local X-ray selected groups, there is evidence for smoother
disks in group galaxies than in the field. Studies of
blue cluster galaxies have also shown that they have significantly lower
asymmetry values than their blue field counterparts
\citep{McIntosh04}. Perhaps most intriguingly, \citet{homeier} has
recently shown that X-ray luminous clusters have galaxies with
significantly lower average asymmetries than X-ray faint clusters. 
A related difference from our results comes from studies of X-ray group
galaxy morphology, which have
shown that the fraction of early types is $\sim$ 0.7 \citep{Mulchaey, Jeltema}
in the same magnitude and redshift range as our CNOC2 sample. 
To compare our data with this number we define early-type galaxies as
those with B/T$>0.4$, as \citet{tran} has shown that this provides a
good match with early type galaxies as classified on the
Hubble-sequence.  Using this definition, we find an early-type fraction
of only $\sim$ 46 $\%$ in the
CNOC2 group
sample, considerably lower than found in X-ray
groups at this redshift.  We will revisit this issue in a future paper
when we consider and compare the results of Hubble-sequence
morphological classification \citep{wilmanMORPH}.
These
results, combined with our result that the median asymmetry in
optically selected group galaxies is not different from the luminosity
matched field, point to the role that the hot intergalactic medium (IGM)
may play in the
smoothing of disk galaxies. Alternatively, the progenitors of
optically selected groups may be different from the progenitors of
X-ray selected ones.

\subsection{Comparison with semi-analytic galaxy models}\label{modresult}
 
Recently, large dark matter simulations of large
volumes have allowed theorists to produce usefully large catalogues of model
galaxies, employing detailed modeling of galaxy formation based
on relatively simple prescriptions for relevant physical processes. In this section
we use the catalogues of one such ``semi-analytic'' galaxy
formation model \citep{bowermodel} to compare with our data. The Bower
et al. model uses the dark matter Millennium simulation
\citep{mil_sim}, a $\Lambda$CDM cosmological box with $500/h$ Mpc
sides, as the basis for the merger trees.  The algorithm is based on the earlier GALFORM models of
\citet{benson_gal} and \citet{cole_gal}. The principal change from the
Benson et al. model is a prescription for the quenching of star
formation in massive halos by feedback powered by accretion onto a supermassive black hole. However, perhaps the most important
change for our purposes is the modification of the method for
computing disk instabilities. Disk instabilities are now the dominant
mode of bulge formation in these models, although the brightest galaxies are
still more often formed through mergers. Unlike older models,
morphology is now sensitive to the baryonic physics of disks, rather
than the (more robustly-predicted) merger history. 
\begin{figure*}
\begin{minipage}{0.45 \linewidth}
\leavevmode \epsfysize=8cm \epsfbox{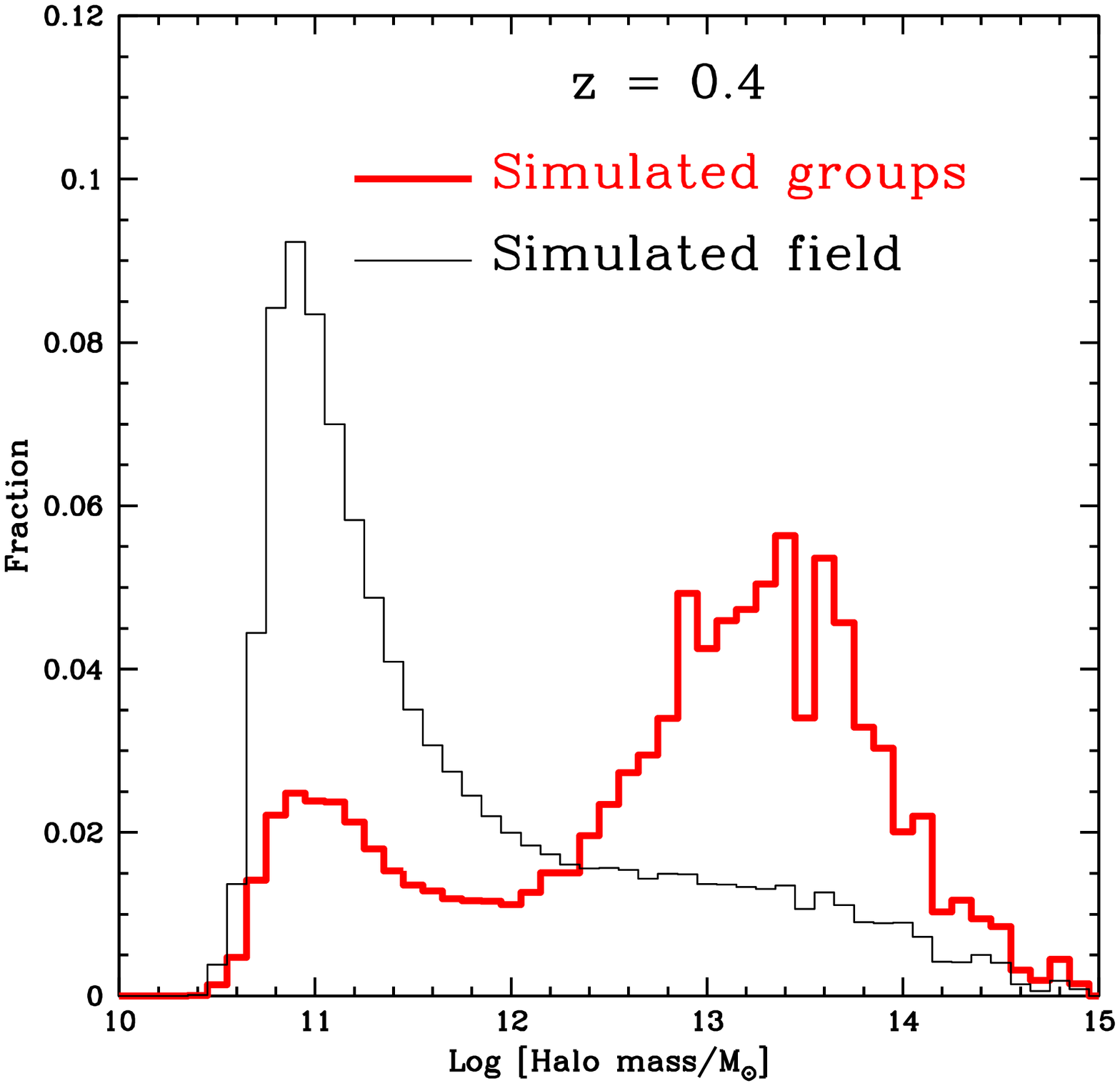}
\end{minipage}
\begin{minipage}{0.45 \linewidth}
\leavevmode \epsfysize=8cm \epsfbox{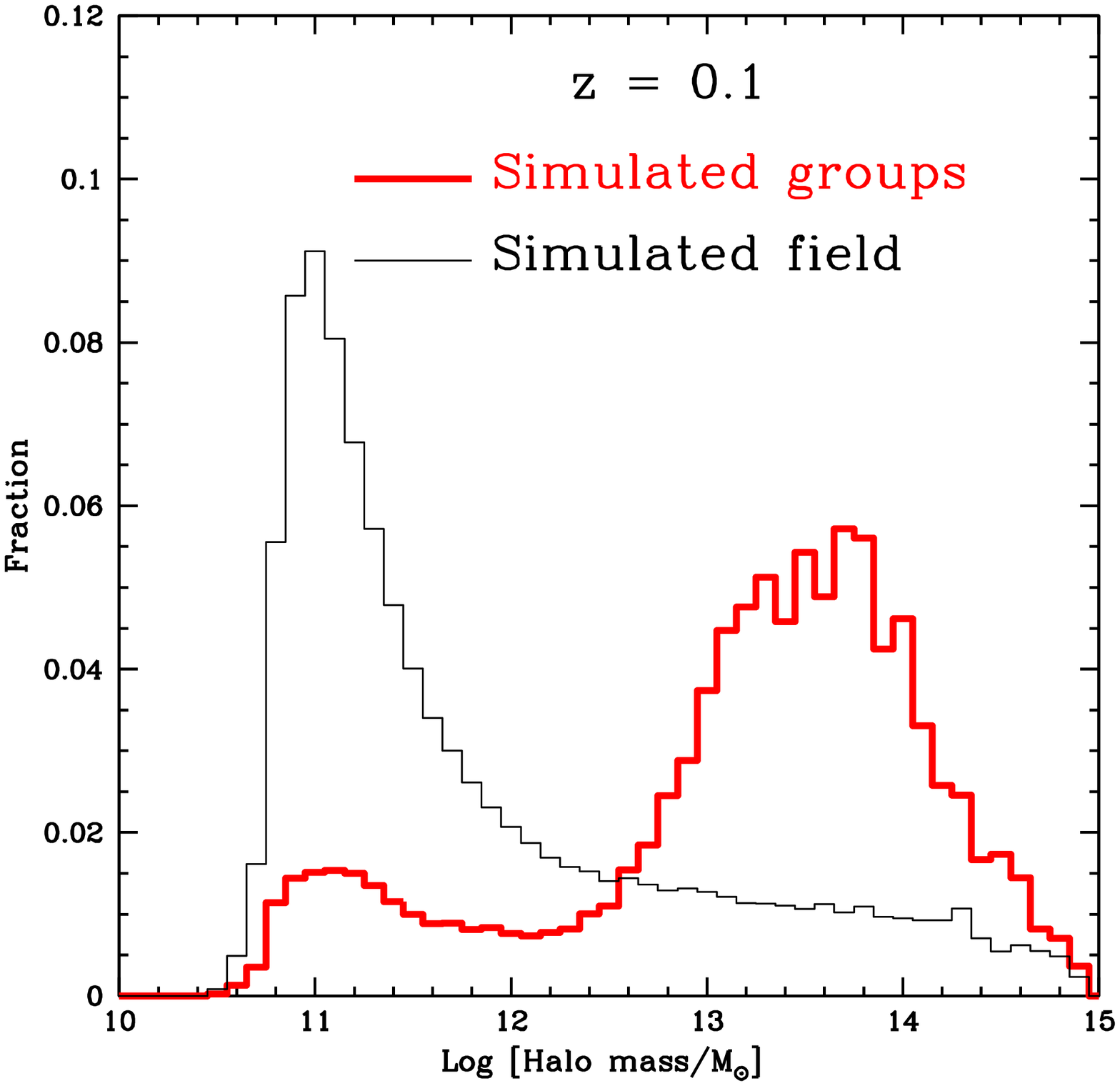}
\end{minipage}
\caption{The distribution of halo masses for each galaxy in the
  \citet{bowermodel} model catalogue (thin, black line) and the distributions of
  galaxy halos in the simulated groups (thick, red line). The high
  redshift sample is on the left and the low redshift sample is on
  the right. }
\label{mil_halos}
\end{figure*} 

\subsubsection{Constructing a mock catalogue} \label{halos}
We construct mock catalogues using the \citet{bowermodel} model for the z=0.1 and z=0.4
redshift time steps. Because of the large size of the Millennium simulation and the small
spread in redshift of each of our samples, a single simulation box can
be used for each epoch.  Using the same group-finding procedure as described in detail
in Appendix \ref{lowfinder}, we construct a stacked group sample to
compare directly with our observations. 

In Figure \ref{mil_halos}, we present the resulting distribution of
galaxy halos in our group sample, compared with all halos in the
Millennium simulation. We see that at both redshifts the group-finding
algorithm selects predominately galaxies that are in halos with masses
$5\times10^{12} < M_{\rm halo}/M_\odot <10^{15}$. Both samples peak
at $M_{\rm halo}\sim$ 6 $\times$ 10$^{13}M_\odot$, but the CNOC2
sample is somewhat biased toward lower masses than the MGC sample. Our
algorithm finds a large percentage of groups that are made
up principally of large dark matter halos: 76.4 $\%$ (85.1 $\%$) of z=0.4 (z=0.1)
galaxies are in halos with  $M_{\rm halo}>10^{12}$
M$_\odot$. In both plots there is a second peak -- a distribution of
low mass halos -- which are contamination. However, we find that 79\%
 (z=0.4) and 89\% (z=0) of the $M_{\rm halo}<$10$^{12}$ M$_\odot$ galaxies are within 500\ho kpc of a
galaxy which resides in a halo with $M_{\rm halo}>$10$^{12}$ M$_\odot$. Thus, the 
majority of our ``contamination'' is due to galaxies on the
outskirts of a true group.  This confirms that the \citet{CNOC2-groups} algorithm selects
groups which are real, and representative of massive dark matter halos, as
also confirmed by previous weak lensing measurements
\citep{parker} and our follow-up spectroscopy \citep{Wilman1}. Only
2.5 $\%$ (3.1 $\%$) of our z=0.4 (z=0.1) galaxies are not associated
with a massive ($M_{\rm halo}>10^{12}M_\odot$) halo.

\begin{figure}
\leavevmode  \epsfysize=8cm  \epsfbox{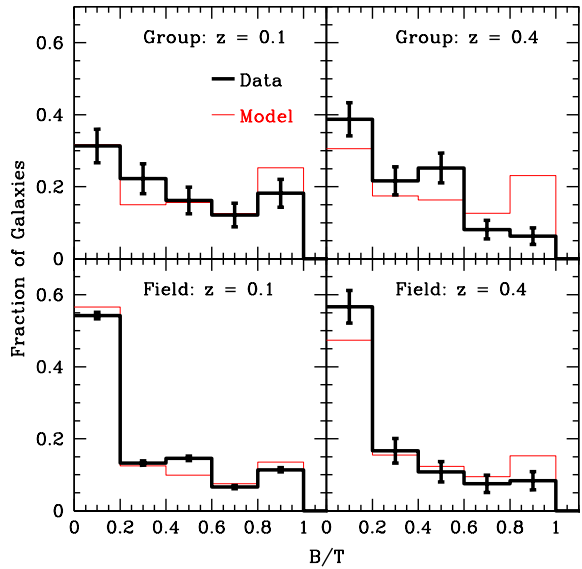}
\caption{The B/T distributions of the \citet{bowermodel} model, compared
  with the observed group and field samples at z=0.1 and z=0.4. The model is the thin, red line in
  each panel and the data is the thick black line.}
\label{bower_bt}
\end{figure}
\subsubsection{B/T distribution}

In Figure \ref{bower_bt}, we show the \citet{bowermodel} model predictions for the
B/T distributions corresponding to our data samples.  In all four panels the data are
shown with a thick black line and the model predictions are shown with a
thin, dashed, red line. The models are limited at $M_B<-18$ in the MGC comparisons, and by $M_v<-19$ for the
CNOC2 comparison.

It is clear that there is remarkable agreement between the model and
the data at z=0.1, especially considering that the model does not take into account any
observational uncertainties associated with deriving B/T ratios from
the surface brightness profile alone. However, the agreement
  between the models and the data at z=0.4 is not as good. In
  particular, the model underpredicts the fraction of B/T $<$ 0.2 galaxies
  and overpredicts the fraction of B/T $>$ 0.8 galaxies in both the
  groups and the field. Intriguingly, Figure \ref{bower_bt} shows the
  models also predict that the fraction of disk dominated galaxies
  increases in the field between z=0.4 and z=0.1, but remains constant
  within the groups. 

\begin{figure}
\leavevmode \epsfysize=8cm \epsfbox{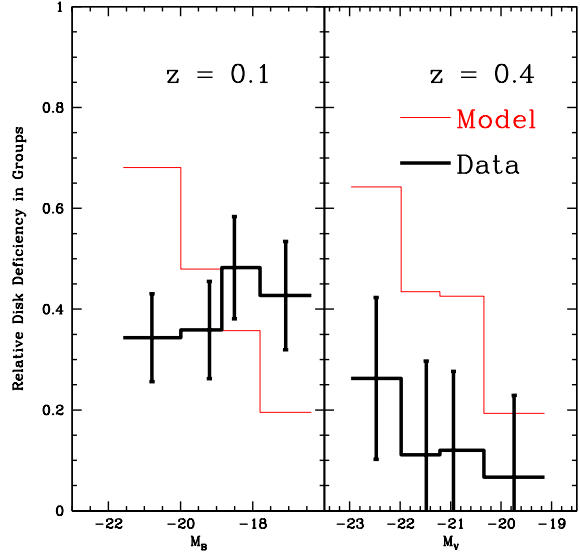}
\caption{The relative disk deficiency in groups, parametrized as
  1-group disk fraction/field disk fraction, where a disk has a B/T
  $<$ 0.2. This is shown at z=0.4 (right panel) and z=0.1 (left panel)
  for both the Bower et al model (thin red line) and the data (thick
  black line).}
\label{bower_deficiency}
\end{figure}

Given these predictions, we are now encouraged to investigate the
  time evolution of the disk fraction as a function of luminosity. From our data we
have seen that the differences in the disk fraction as a function of
magnitude are small at z=0.4 ($\sim$ 5.5 $\pm$ 2 $\%$), but quite
large in the local universe ($\sim$ 19 $\pm$ 6 $\%$). 
To address this, we present Figure \ref{bower_deficiency}, which shows the
relative disk (B/T $<$ 0.2) deficiency between the field and the group samples, as a function of
luminosity. The disk deficiency is the difference between the group
and field disk fraction divided by the field disk fraction. This gives
a measure of the fraction of field disks which are absent at a similar
magnitude in the groups. In the left panel of Figure \ref{bower_deficiency} we show the low redshift
sample. The data from the MGC sample agrees well on average with the
model predictions. In the right panel of Figure \ref{bower_deficiency} we show the same
comparison but for the high redshift sample. Although the average
value of the group disk deficit is correctly predicted at z=0.1, the
Bower et al. model predicts a group disk deficit which is much higher
than the data at z=0.4. We note that, in the model, the
predicted evolution is of a similar magnitude whether measured
consistently in rest frame V or rest frame B, indirectly supporting
our argument that the observed evolution is not driven by the
difference in rest wavelength sampled by the two surveys.

We have seen in Figure \ref{bower_bt} that the models
underpredict the fraction of B/T $<$ 0.2 at z=0.4 and that this leads
to an overprediction of the disk deficiency in groups at the high
redshift epoch.  Intriguingly, inspection of Figure \ref{bower_bt}
shows that this is because the models predict that the fraction of disk dominated galaxies
  increases in the field between z=0.4 and z=0.1, but remains constant
  within the groups. While a direct comparison between the two epochs
  could be complicated by the different observed wavebands, we note that
  the same model predictions exist when considering only the
  underlying stellar mass. This is at odds with the data, which show
  that the B/T $<$ 0.2 fraction {\it decreases} in the groups, and
  remains constant in the field.

\section{Conclusions}\label{summary}
We have presented a quantitative morphology study of
optically-selected galaxy groups 
from two redshift surveys: CNOC2 (z$\sim$0.4), supplemented 
with significant additional Magellan spectroscopy, and MGC (z$\sim$0.1). We
have compared these data with a similarly selected sample of groups
drawn from the semi-analytic galaxy formation models of
\citet{bowermodel}. Our findings are:

\begin{itemize}
 
\item There is a significant difference, as indicated by a KS test,
  in the fractional bulge luminosity (B/T) distribution of group and field galaxies, in both the high and
  low redshift samples. The dominant difference is the deficit of
  disk--dominated (B/T $<$ 0.2) galaxies in the group samples. 

\item The difference in the disk fraction (B/T$<$0.2) of group
  galaxies relative to the field
  shows significant evolution between z=0.4 and z=0.1. At a given luminosity in the CNOC2 sample, the groups have $\sim$ 5.5
  $\pm$ 2 $\%$ fewer B/T$<$0.2 galaxies than the field. By z=0.1 this
  difference has increased significantly, so that groups have
  $\sim$ 19 $\pm$ 6 $\%$ fewer B/T$<$0.2 galaxies than the field in
  the same magnitude range. Although the z=0.1 sample traces rest
  frame B while the z=0.4 sample traces rest frame V, this difference is
  unlikely to be able to explain the differences. 

\item At neither redshift do we see any evidence that the bulk
  properties of the existing disks are significantly different for
  group galaxies than for field galaxies. They lie on similar scaling
  relations and show similar asymmetry distributions. There is no evidence
  that groups are actively perturbing or otherwise affecting a large
  fraction of the group disk population. 

\item We find that
  there is a small enhancement in the fraction of bulge-dominated
  group galaxies that are highly asymmetric, relative to
  bulge-dominated galaxies in the field. This may be consistent with
  enhanced merging in the group environment. Visual inspection of
  high asymmetry, bulge-dominated CNOC2 galaxies shows that 44$\%$
  (4/9) exhibit clear evidence of interactions or merging.  

\item A sample of galaxies drawn from the semi-analytic galaxy
  catalogues of \citet{bowermodel} was shown to agree remarkably well
  with the B/T distribution of the field and group galaxies at
  z=0.1. However, our data have shown that time evolution of the B/T
  distributions predicted by the models is not seen in our data. In
  particular, the Bower et al. model underpredicts the fraction of
  disk dominated galaxies at z $\sim$ 0.4.

\end{itemize}

The morphological difference
between group and field galaxies at z=0.4 is mostly due to the tendency for
group galaxies to be more luminous and, therefore, more bulge-dominated than field galaxies. This is
consistent with our previous findings about group galaxies; namely, that
their M/L ratios are consistent with the passive evolution of a
predominately old population \citep{Balogh_smass}, because the
dominant difference in group galaxies is their pre-existing tendency to
be bulge-dominated. This is also consistent with the fairly small difference 
in the emission-line fraction of group and field galaxies, at fixed stellar mass 
\citep{Balogh_smass} or magnitude \citep{Wilman2}.  

The failure of the \citet{bowermodel} to reproduce the time evolution
of the group disk deficit is interesting. 
These type of models predict a fairly rapid "strangulation" of galaxies once they enter
larger halos, which causes the star formation rate to decrease on
timescales of a few Gyr.  This starts to play a large role at $\sim$10$^{12}$
solar masses, significantly less massive than most of our groups.  
\citet{weinmann1} have shown that this mechanism is too effective, and
produces a homogeneous, red satellite population at all magnitudes, in
groups and clusters, which is not observed at z=0. \citet{Gilbank}
have recently shown that this problem extends out to at least z=1. It
is likely that the incorrect disk deficit evolution is another
manifestation of the maximally efficient star formation quenching
mechanism used by the models.

We have found that, in contrast to
X-ray selected groups, our optically-selected group galaxies have median
asymmetries that are similar to field
galaxies.  
This may point to a possible role for galaxy
interactions with the hot IGM. Alternatively, the progenitors of X-ray selected group
galaxies may be 
fundamentally different from the progenitors of optically selected
group galaxies. X-ray selected groups are more likely to be relaxed, virialized
structures, suggesting that they were assembled earlier than optically
selected groups.  In the current models, the strangulation mechanism
is assumed to operate efficiently in small haloes.  Thus, little
difference is expected in the population of different types of groups
many Gyr later (i.e. at the epochs of interest here), since star
formation has long ceased in most group members.  However, recently it has been suggested that infalling
galaxies may be able to retain a significant fraction of their gas
\citep{mccarthy-ram, kawata},
significantly increasing the timescale for star formation to decrease.  In this case, the
headstart given to galaxies that fall into groups a little earlier may be better
able to
explain the difference between X-ray and optically selected groups.

\section*{Acknowledgments}
We thank the CNOC2 team for access to their unpublished data and Luc
Simard for making \textsc{gim2d} publicly available. We also thank
the GALFORM team for allowing access to the semi-analytic galaxy
catalogues. Finally, we thank the MGC team for their data. 
The Millennium Galaxy Catalogue consists of imaging data from the
Isaac Newton Telescope and spectroscopic data from the Anglo
Australian Telescope, the ANU 2.3m, the ESO New Technology Telescope,
the Telescopio Nazionale Galileo and the Gemini North Telescope. The
survey has been supported through grants from the Particle Physics and
Astronomy Research Council (UK) and the Australian Research Council
(AUS). The data and data products are publicly available from
http://www.eso.org/$\sim$jliske/mgc/ or on request from J. Liske or
S.P. Driver.  MLB acknowledges support from an NSERC Discovery Grant.

\bibliography{ms}

\begin{thebibliography}{67}
\expandafter\ifx\csname natexlab\endcsname\relax\def\natexlab#1{#1}\fi

\bibitem[{{Abraham} {et~al.}(1991){Abraham}, {Crawford}, \&
  {McHardy}}]{91abraham}
{Abraham}, R.~G., {Crawford}, C.~S., \& {McHardy}, I.~M. 1991, MNRAS, 252, 482

\bibitem[{{Abraham} {et~al.}(2003){Abraham}, {van den Bergh}, \&
  {Nair}}]{abraham03}
{Abraham}, R.~G., {van den Bergh}, S., \& {Nair}, P. 2003, ApJ, 588, 218

\bibitem[{{Allen} {et~al.}(2006){Allen}, {Driver}, {Graham}, {Cameron},
  {Liske}, \& {de Propris}}]{mgc-gim2d}
{Allen}, P.~D., {Driver}, S.~P., {Graham}, A.~W., {Cameron}, E., {Liske}, J.,
  \& {de Propris}, R. 2006, MNRAS, 371, 2

\bibitem[{{Baldry} {et~al.}(2004){Baldry}, {Glazebrook}, {Brinkmann},
  {Ivezi{\'c}}, {Lupton}, {Nichol}, \& {Szalay}}]{baldry}
{Baldry}, I.~K., {Glazebrook}, K., {Brinkmann}, J., {Ivezi{\'c}}, {\v Z}.,
  {Lupton}, R.~H., {Nichol}, R.~C., \& {Szalay}, A.~S. 2004, apj, 600, 681

\bibitem[{{Balogh} {et~al.}(2004){Balogh}, {Eke}, {Miller}, {Lewis}, {Bower},
  {Couch}, {Nichol}, {Bland-Hawthorn}, {Baldry}, {Baugh}, {Bridges}, {Cannon},
  {Cole}, {Colless}, {Collins}, {Cross}, {Dalton}, {de Propris}, {Driver},
  {Efstathiou}, {Ellis}, {Frenk}, {Glazebrook}, {Gomez}, {Gray}, {Hawkins},
  {Jackson}, {Lahav}, {Lumsden}, {Maddox}, {Madgwick}, {Norberg}, {Peacock},
  {Percival}, {Peterson}, {Sutherland}, \& {Taylor}}]{balogh_ecology}
{Balogh}, M., {Eke}, V., {Miller}, C., {Lewis}, I., {Bower}, R., {Couch}, W.,
  {Nichol}, R., {Bland-Hawthorn}, J., {Baldry}, I.~K., {Baugh}, C., {Bridges},
  T., {Cannon}, R., {Cole}, S., {Colless}, M., {Collins}, C., {Cross}, N.,
  {Dalton}, G., {de Propris}, R., {Driver}, S.~P., {Efstathiou}, G., {Ellis},
  R.~S., {Frenk}, C.~S., {Glazebrook}, K., {Gomez}, P., {Gray}, A., {Hawkins},
  E., {Jackson}, C., {Lahav}, O., {Lumsden}, S., {Maddox}, S., {Madgwick}, D.,
  {Norberg}, P., {Peacock}, J.~A., {Percival}, W., {Peterson}, B.~A.,
  {Sutherland}, W., \& {Taylor}, K. 2004, MNRAS, 348, 1355

\bibitem[{{Balogh} {et~al.}(1999){Balogh}, {Morris}, {Yee}, {Carlberg}, \&
  {Ellingson}}]{balogh}
{Balogh}, M.~L., {Morris}, S.~L., {Yee}, H.~K.~C., {Carlberg}, R.~G., \&
  {Ellingson}, E. 1999, ApJ, 527, 54

\bibitem[{{Balogh} {et~al.}(2007){Balogh}, {Wilman}, {Henderson}, {Bower},
  {Gilbank}, {Whitaker}, {Morris}, {Hau}, {Mulchaey}, {Oemler}, \&
  {Carlberg}}]{Balogh_smass}
{Balogh}, M.~L., {Wilman}, D., {Henderson}, R.~D.~E., {Bower}, R.~G.,
  {Gilbank}, D., {Whitaker}, R., {Morris}, S.~L., {Hau}, G., {Mulchaey}, J.~S.,
  {Oemler}, A., \& {Carlberg}, R.~G. 2007, MNRAS, 374, 1169

\bibitem[{{Beers} {et~al.}(1990){Beers}, {Flynn}, \& {Gebhardt}}]{Gapper}
{Beers}, T.~C., {Flynn}, K., \& {Gebhardt}, K. 1990, AJ, 100, 32

\bibitem[{{Bell} {et~al.}(2004){Bell}, {Wolf}, {Meisenheimer}, {Rix}, {Borch},
  {Dye}, {Kleinheinrich}, {Wisotzki}, \& {McIntosh}}]{bell}
{Bell}, E.~F., {Wolf}, C., {Meisenheimer}, K., {Rix}, H.-W., {Borch}, A.,
  {Dye}, S., {Kleinheinrich}, M., {Wisotzki}, L., \& {McIntosh}, D.~H. 2004,
  ApJ, 608, 752

\bibitem[{{Bell} {et~al.}(2007){Bell}, {Zheng}, {Papovich}, {Borch}, {Wolf}, \&
  {Meisenheimer}}]{Bell07}
{Bell}, E.~F., {Zheng}, X.~Z., {Papovich}, C., {Borch}, A., {Wolf}, C., \&
  {Meisenheimer}, K. 2007, ApJ, 663, 834

\bibitem[{{Benson} {et~al.}(2003){Benson}, {Bower}, {Frenk}, {Lacey}, {Baugh},
  \& {Cole}}]{benson_gal}
{Benson}, A.~J., {Bower}, R.~G., {Frenk}, C.~S., {Lacey}, C.~G., {Baugh},
  C.~M., \& {Cole}, S. 2003, ApJ, 599, 38

\bibitem[{{Berlind} {et~al.}(2006){Berlind}, {Frieman}, {Weinberg}, {Blanton},
  {Warren}, {Abazajian}, {Scranton}, {Hogg}, {Scoccimarro}, {Bahcall},
  {Brinkmann}, {Gott}, {Kleinman}, {Krzesinski}, {Lee}, {Miller}, {Nitta},
  {Schneider}, {Tucker}, \& {Zehavi}}]{Berlind}
{Berlind}, A.~A., {Frieman}, J., {Weinberg}, D.~H., {Blanton}, M.~R., {Warren},
  M.~S., {Abazajian}, K., {Scranton}, R., {Hogg}, D.~W., {Scoccimarro}, R.,
  {Bahcall}, N.~A., {Brinkmann}, J., {Gott}, J.~R.~I., {Kleinman}, S.~J.,
  {Krzesinski}, J., {Lee}, B.~C., {Miller}, C.~J., {Nitta}, A., {Schneider},
  D.~P., {Tucker}, D.~L., \& {Zehavi}, I. 2006, ApJS, 167, 1

\bibitem[{Bertin \& Arnouts(1996)}]{sextractor}
Bertin, E. \& Arnouts, S. 1996, A\&AS, 117, 393

\bibitem[{{Blanton} {et~al.}(2003){Blanton}, {Hogg}, {Bahcall}, {Baldry},
  {Brinkmann}, {Csabai}, {Eisenstein}, {Fukugita}, {Gunn}, {Ivezi{\'c}},
  {Lamb}, {Lupton}, {Loveday}, {Munn}, {Nichol}, {Okamura}, {Schlegel},
  {Shimasaku}, {Strauss}, {Vogeley}, \& {Weinberg}}]{Blanton03}
{Blanton}, M.~R., {Hogg}, D.~W., {Bahcall}, N.~A., {Baldry}, I.~K.,
  {Brinkmann}, J., {Csabai}, I., {Eisenstein}, D., {Fukugita}, M., {Gunn},
  J.~E., {Ivezi{\'c}}, {\v Z}., {Lamb}, D.~Q., {Lupton}, R.~H., {Loveday}, J.,
  {Munn}, J.~A., {Nichol}, R.~C., {Okamura}, S., {Schlegel}, D.~J.,
  {Shimasaku}, K., {Strauss}, M.~A., {Vogeley}, M.~S., \& {Weinberg}, D.~H.
  2003, ApJ, 594, 186

\bibitem[{{Bower} {et~al.}(2006){Bower}, {Benson}, {Malbon}, {Helly}, {Frenk},
  {Baugh}, {Cole}, \& {Lacey}}]{bowermodel}
{Bower}, R.~G., {Benson}, A.~J., {Malbon}, R., {Helly}, J.~C., {Frenk}, C.~S.,
  {Baugh}, C.~M., {Cole}, S., \& {Lacey}, C.~G. 2006, MNRAS, 370, 645

\bibitem[{{Bruzual} \& {Charlot}(2003)}]{bc03}
{Bruzual}, G. \& {Charlot}, S. 2003, MNRAS, 344, 1000

\bibitem[{{Carlberg} {et~al.}(2001){Carlberg}, {Yee}, {Morris}, {Lin}, {Hall},
  {Patton}, {Sawicki}, \& {Shepherd}}]{CNOC2-groups}
{Carlberg}, R.~G., {Yee}, H.~K.~C., {Morris}, S.~L., {Lin}, H., {Hall}, P.~B.,
  {Patton}, D.~R., {Sawicki}, M., \& {Shepherd}, C.~W. 2001, ApJ, 552, 427

\bibitem[{{Cole} {et~al.}(2000){Cole}, {Lacey}, {Baugh}, \& {Frenk}}]{cole_gal}
{Cole}, S., {Lacey}, C.~G., {Baugh}, C.~M., \& {Frenk}, C.~S. 2000, MNRAS, 319,
  168

\bibitem[{{Croton} {et~al.}(2006){Croton}, {Springel}, {White}, {De Lucia},
  {Frenk}, {Gao}, {Jenkins}, {Kauffmann}, {Navarro}, \& {Yoshida}}]{croton}
{Croton}, D.~J., {Springel}, V., {White}, S.~D.~M., {De Lucia}, G., {Frenk},
  C.~S., {Gao}, L., {Jenkins}, A., {Kauffmann}, G., {Navarro}, J.~F., \&
  {Yoshida}, N. 2006, MNRAS, 365, 11

\bibitem[{{Dressler}(1980)}]{Dressler80}
{Dressler}, A. 1980, ApJ, 236, 351

\bibitem[{{Dressler} {et~al.}(1997){Dressler}, {Oemler}, {Couch}, {Smail},
  {Ellis}, {Barger}, {Butcher}, {Poggianti}, \& {Sharples}}]{Dressler97}
{Dressler}, A., {Oemler}, A.~J., {Couch}, W.~J., {Smail}, I., {Ellis}, R.~S.,
  {Barger}, A., {Butcher}, H., {Poggianti}, B.~M., \& {Sharples}, R.~M. 1997,
  ApJ, 490, 577

\bibitem[{{Driver} {et~al.}(2005){Driver}, {Liske}, {Cross}, {De Propris}, \&
  {Allen}}]{MGC-spectra}
{Driver}, S.~P., {Liske}, J., {Cross}, N.~J.~G., {De Propris}, R., \& {Allen},
  P.~D. 2005, MNRAS, 360, 81

\bibitem[{{Eke} {et~al.}(2004){Eke}, {Baugh}, {Cole}, {Frenk}, {Norberg},
  {Peacock}, {Baldry}, {Bland-Hawthorn}, {Bridges}, {Cannon}, {Colless},
  {Collins}, {Couch}, {Dalton}, {de Propris}, {Driver}, {Efstathiou}, {Ellis},
  {Glazebrook}, {Jackson}, {Lahav}, {Lewis}, {Lumsden}, {Maddox}, {Madgwick},
  {Peterson}, {Sutherland}, \& {Taylor}}]{2PIGG-cat}
{Eke}, V.~R., {Baugh}, C.~M., {Cole}, S., {Frenk}, C.~S., {Norberg}, P.,
  {Peacock}, J.~A., {Baldry}, I.~K., {Bland-Hawthorn}, J., {Bridges}, T.,
  {Cannon}, R., {Colless}, M., {Collins}, C., {Couch}, W., {Dalton}, G., {de
  Propris}, R., {Driver}, S.~P., {Efstathiou}, G., {Ellis}, R.~S.,
  {Glazebrook}, K., {Jackson}, C., {Lahav}, O., {Lewis}, I., {Lumsden}, S.,
  {Maddox}, S., {Madgwick}, D., {Peterson}, B.~A., {Sutherland}, W., \&
  {Taylor}, K. 2004, MNRAS, 348, 866

\bibitem[{{Faber} {et~al.}(2007){Faber}, {Willmer}, {Wolf}, {Koo}, {Weiner},
  {Newman}, {Im}, {Coil}, {Conroy}, {Cooper}, {Davis}, {Finkbeiner}, {Gerke},
  {Gebhardt}, {Groth}, {Guhathakurta}, {Harker}, {Kaiser}, {Kassin},
  {Kleinheinrich}, {Konidaris}, {Kron}, {Lin}, {Luppino}, {Madgwick},
  {Meisenheimer}, {Noeske}, {Phillips}, {Sarajedini}, {Schiavon}, {Simard},
  {Szalay}, {Vogt}, \& {Yan}}]{faber}
{Faber}, S.~M., {Willmer}, C.~N.~A., {Wolf}, C., {Koo}, D.~C., {Weiner}, B.~J.,
  {Newman}, J.~A., {Im}, M., {Coil}, A.~L., {Conroy}, C., {Cooper}, M.~C.,
  {Davis}, M., {Finkbeiner}, D.~P., {Gerke}, B.~F., {Gebhardt}, K., {Groth},
  E.~J., {Guhathakurta}, P., {Harker}, J., {Kaiser}, N., {Kassin}, S.,
  {Kleinheinrich}, M., {Konidaris}, N.~P., {Kron}, R.~G., {Lin}, L., {Luppino},
  G., {Madgwick}, D.~S., {Meisenheimer}, K., {Noeske}, K.~G., {Phillips},
  A.~C., {Sarajedini}, V.~L., {Schiavon}, R.~P., {Simard}, L., {Szalay}, A.~S.,
  {Vogt}, N.~P., \& {Yan}, R. 2007, ApJ, 665, 265

\bibitem[{{Fabian} {et~al.}(2000){Fabian}, {Sanders}, {Ettori}, {Taylor},
  {Allen}, {Crawford}, {Iwasawa}, {Johnstone}, \& {Ogle}}]{Fabian00}
{Fabian}, A.~C., {Sanders}, J.~S., {Ettori}, S., {Taylor}, G.~B., {Allen},
  S.~W., {Crawford}, C.~S., {Iwasawa}, K., {Johnstone}, R.~M., \& {Ogle}, P.~M.
  2000, MNRAS, 318, L65

\bibitem[{{Ferrarese} \& {Merritt}(2000)}]{Ferrarese00}
{Ferrarese}, L. \& {Merritt}, D. 2000, ApJL, 539, L9

\bibitem[{{Fukugita} {et~al.}(1995){Fukugita}, {Shimasaku}, \&
  {Ichikawa}}]{fukugita}
{Fukugita}, M., {Shimasaku}, K., \& {Ichikawa}, T. 1995, PASP, 107, 945

\bibitem[{{Gilbank} \& {Balogh}(2008)}]{Gilbank}
{Gilbank}, D.~G. \& {Balogh}, M.~L. 2008, mnras, L18+

\bibitem[{{H{\"a}ussler} {et~al.}(2007){H{\"a}ussler}, {McIntosh}, {Barden},
  {Bell}, {Rix}, {Borch}, {Beckwith}, {Caldwell}, {Heymans}, {Jahnke}, {Jogee},
  {Koposov}, {Meisenheimer}, {S{\'a}nchez}, {Somerville}, {Wisotzki}, \&
  {Wolf}}]{haussler}
{H{\"a}ussler}, B., {McIntosh}, D.~H., {Barden}, M., {Bell}, E.~F., {Rix},
  H.-W., {Borch}, A., {Beckwith}, S.~V.~W., {Caldwell}, J.~A.~R., {Heymans},
  C., {Jahnke}, K., {Jogee}, S., {Koposov}, S.~E., {Meisenheimer}, K.,
  {S{\'a}nchez}, S.~F., {Somerville}, R.~S., {Wisotzki}, L., \& {Wolf}, C.
  2007, apjs, 172, 615

\bibitem[{{Homeier} {et~al.}(2006){Homeier}, {Postman}, {Menanteau},
  {Blakeslee}, {Mei}, {Demarco}, {Ford}, {Illingworth}, \& {Zirm}}]{homeier}
{Homeier}, N.~L., {Postman}, M., {Menanteau}, F., {Blakeslee}, J.~P., {Mei},
  S., {Demarco}, R., {Ford}, H.~C., {Illingworth}, G.~D., \& {Zirm}, A. 2006,
  AJ, 131, 143

\bibitem[{{Hopkins}(2004)}]{hopkins2004}
{Hopkins}, A.~M. 2004, ApJ, 615, 209

\bibitem[{{Hopkins} {et~al.}(2007){Hopkins}, {Bundy}, {Hernquist}, \&
  {Ellis}}]{Hopkins07}
{Hopkins}, P.~F., {Bundy}, K., {Hernquist}, L., \& {Ellis}, R.~S. 2007, ApJ,
  659, 976

\bibitem[{Im {et~al.}(2002)Im, Simard, Faber, Koo, Gebhardt, Willmer, Phillips,
  Illingworth, Vogt, \& Sarajedini}]{Grothx}
Im, M., Simard, L., Faber, S.~M., Koo, D.~C., Gebhardt, K., Willmer, C. N.~A.,
  Phillips, A.~C., Illingworth, G.~D., Vogt, N.~P., \& Sarajedini, V.~L. 2002,
  ApJ, 571, 136

\bibitem[{{Jeltema} {et~al.}(2007){Jeltema}, {Mulchaey}, {Lubin}, \&
  {Fassnacht}}]{Jeltema}
{Jeltema}, T.~E., {Mulchaey}, J.~S., {Lubin}, L.~M., \& {Fassnacht}, C.~D.
  2007, ApJ, 658, 865

\bibitem[{{Kawata} \& {Mulchaey}(2007)}]{kawata}
{Kawata}, D. \& {Mulchaey}, J.~S. 2007, ArXiv e-prints, 707

\bibitem[{{Lilly} {et~al.}(1996){Lilly}, {Le Fevre}, {Hammer}, \&
  {Crampton}}]{lilly96}
{Lilly}, S.~J., {Le Fevre}, O., {Hammer}, F., \& {Crampton}, D. 1996, ApJL,
  460, L1+

\bibitem[{{Liske} {et~al.}(2003){Liske}, {Lemon}, {Driver}, {Cross}, \&
  {Couch}}]{MGC-phot}
{Liske}, J., {Lemon}, D.~J., {Driver}, S.~P., {Cross}, N.~J.~G., \& {Couch},
  W.~J. 2003, MNRAS, 344, 307

\bibitem[{{Lotz} {et~al.}(2004){Lotz}, {Primack}, \& {Madau}}]{lotz}
{Lotz}, J.~M., {Primack}, J., \& {Madau}, P. 2004, AJ, 128, 163

\bibitem[{{Madau} {et~al.}(1996){Madau}, {Ferguson}, {Dickinson}, {Giavalisco},
  {Steidel}, \& {Fruchter}}]{madau96}
{Madau}, P., {Ferguson}, H.~C., {Dickinson}, M.~E., {Giavalisco}, M.,
  {Steidel}, C.~C., \& {Fruchter}, A. 1996, MNRAS, 283, 1388

\bibitem[{{Magorrian} {et~al.}(1998){Magorrian}, {Tremaine}, {Richstone},
  {Bender}, {Bower}, {Dressler}, {Faber}, {Gebhardt}, {Green}, {Grillmair},
  {Kormendy}, \& {Lauer}}]{Magorrian97}
{Magorrian}, J., {Tremaine}, S., {Richstone}, D., {Bender}, R., {Bower}, G.,
  {Dressler}, A., {Faber}, S.~M., {Gebhardt}, K., {Green}, R., {Grillmair}, C.,
  {Kormendy}, J., \& {Lauer}, T. 1998, AJ, 115, 2285

\bibitem[{{Marleau} \& {Simard}(1998)}]{MS}
{Marleau}, F.~R. \& {Simard}, L. 1998, ApJ, 507, 585

\bibitem[{{McCarthy} {et~al.}(2007){McCarthy}, {Frenk}, {Font}, {Lacey},
  {Bower}, {Mitchell}, {Balogh}, \& {Theuns}}]{mccarthy-ram}
{McCarthy}, I.~G., {Frenk}, C.~S., {Font}, A.~S., {Lacey}, C.~G., {Bower},
  R.~G., {Mitchell}, N.~L., {Balogh}, M.~L., \& {Theuns}, T. 2007, ArXiv
  e-prints, 710

\bibitem[{{McIntosh} {et~al.}(2004){McIntosh}, {Rix}, \&
  {Caldwell}}]{McIntosh04}
{McIntosh}, D., {Rix}, H.-W., \& {Caldwell}, N. 2004, ApJ, 610, 161

\bibitem[{{McNamara} {et~al.}(2000){McNamara}, {Wise}, {Nulsen}, {David},
  {Sarazin}, {Bautz}, {Markevitch}, {Vikhlinin}, {Forman}, {Jones}, \&
  {Harris}}]{McnamaraHydraA}
{McNamara}, B.~R., {Wise}, M., {Nulsen}, P.~E.~J., {David}, L.~P., {Sarazin},
  C.~L., {Bautz}, M., {Markevitch}, M., {Vikhlinin}, A., {Forman}, W.~R.,
  {Jones}, C., \& {Harris}, D.~E. 2000, ApJL, 534, L135

\bibitem[{{Menanteau} {et~al.}(2006){Menanteau}, {Ford}, {Motta},
  {Ben{\'{\i}}tez}, {Martel}, {Blakeslee}, \& {Infante}}]{menanteau}
{Menanteau}, F., {Ford}, H.~C., {Motta}, V., {Ben{\'{\i}}tez}, N., {Martel},
  A.~R., {Blakeslee}, J.~P., \& {Infante}, L. 2006, AJ, 131, 208

\bibitem[{Metropolis {et~al.}(1953)Metropolis, Rosenbluth, Rosenbluth, Teller,
  \& Teller}]{metro}
Metropolis, N., Rosenbluth, A., Rosenbluth, M., Teller, A., \& Teller, E. 1953,
  Journal of Chemical Physics, 21, 1087

\bibitem[{{Mihos} \& {Hernquist}(1996)}]{MihosHern}
{Mihos}, J.~C. \& {Hernquist}, L. 1996, ApJ, 464, 641

\bibitem[{{Mulchaey} {et~al.}(2006){Mulchaey}, {Lubin}, {Fassnacht}, {Rosati},
  \& {Jeltema}}]{Mulchaey}
{Mulchaey}, J.~S., {Lubin}, L.~M., {Fassnacht}, C., {Rosati}, P., \& {Jeltema},
  T.~E. 2006, ApJ, 646, 133

\bibitem[{{Parker} {et~al.}(2005){Parker}, {Hudson}, {Carlberg}, \&
  {Hoekstra}}]{parker}
{Parker}, L.~C., {Hudson}, M.~J., {Carlberg}, R.~G., \& {Hoekstra}, H. 2005,
  ApJ, 634, 806

\bibitem[{{Patton} {et~al.}(2005){Patton}, {Grant}, {Simard}, {Pritchet},
  {Carlberg}, \& {Borne}}]{pattonasym}
{Patton}, D.~R., {Grant}, J.~K., {Simard}, L., {Pritchet}, C.~J., {Carlberg},
  R.~G., \& {Borne}, K.~D. 2005, AJ, 130, 2043

\bibitem[{{Pavlovsky et. al}(2005)}]{ACS-pipe}
{Pavlovsky et. al}, C. 2005, ACS Data Handbook, Version 4.0

\bibitem[{{Peng} {et~al.}(2002){Peng}, {Ho}, {Impey}, \& {Rix}}]{galfit}
{Peng}, C.~Y., {Ho}, L.~C., {Impey}, C.~D., \& {Rix}, H.-W. 2002, AJ, 124, 266

\bibitem[{{Poggianti} {et~al.}(1999){Poggianti}, {Smail}, {Dressler}, {Couch},
  {Barger}, {Butcher}, {Ellis}, \& {Oemler}}]{pog_smail}
{Poggianti}, B.~M., {Smail}, I., {Dressler}, A., {Couch}, W.~J., {Barger},
  A.~J., {Butcher}, H., {Ellis}, R.~S., \& {Oemler}, A.~J. 1999, ApJ, 518, 576

\bibitem[{{Postman} \& {Geller}(1984)}]{Postman}
{Postman}, M. \& {Geller}, M.~J. 1984, ApJ, 281, 95

\bibitem[{Schade {et~al.}(1995)Schade, Lilly, Crampton, LeF\`{e}vre, Hammer, \&
  Tresse}]{Schade1}
Schade, D., Lilly, S.~J., Crampton, D., LeF\`{e}vre, O., Hammer, F., \& Tresse,
  L. 1995, ApJ, 451, 1

\bibitem[{Simard {et~al.}(2002)Simard, Willmer, Vogt, Sarajedini, Phillips,
  Weiner, Koo, Im, Illingworth, \& Faber}]{Gim2d}
Simard, L., Willmer, C. N.~A., Vogt, N.~P., Sarajedini, V.~L., Phillips, A.~C.,
  Weiner, B.~J., Koo, D.~C., Im, M., Illingworth, G.~D., \& Faber, S.~M. 2002,
  ApJS, 142, 1

\bibitem[{{Springel} {et~al.}(2005){Springel}, {White}, {Jenkins}, {Frenk},
  {Yoshida}, {Gao}, {Navarro}, {Thacker}, {Croton}, {Helly}, {Peacock}, {Cole},
  {Thomas}, {Couchman}, {Evrard}, {Colberg}, \& {Pearce}}]{mil_sim}
{Springel}, V., {White}, S.~D.~M., {Jenkins}, A., {Frenk}, C.~S., {Yoshida},
  N., {Gao}, L., {Navarro}, J., {Thacker}, R., {Croton}, D., {Helly}, J.,
  {Peacock}, J.~A., {Cole}, S., {Thomas}, P., {Couchman}, H., {Evrard}, A.,
  {Colberg}, J., \& {Pearce}, F. 2005, Nature, 435, 629

\bibitem[{{Taylor-Mager} {et~al.}(2007){Taylor-Mager}, {Conselice},
  {Windhorst}, \& {Jansen}}]{taylor}
{Taylor-Mager}, V.~A., {Conselice}, C.~J., {Windhorst}, R.~A., \& {Jansen},
  R.~A. 2007, ApJ, 659, 162

\bibitem[{{Toomre} \& {Toomre}(1972)}]{ToomreToomre}
{Toomre}, A. \& {Toomre}, J. 1972, ApJ, 178, 623

\bibitem[{{Tran} {et~al.}(2001){Tran}, {Simard}, {Zabludoff}, \&
  {Mulchaey}}]{tran}
{Tran}, K.-V.~H., {Simard}, L., {Zabludoff}, A.~I., \& {Mulchaey}, J.~S. 2001,
  ApJ, 549, 172

\bibitem[{{Weinmann} {et~al.}(2006){Weinmann}, {van den Bosch}, {Yang}, \&
  {Mo}}]{weinmann1}
{Weinmann}, S.~M., {van den Bosch}, F.~C., {Yang}, X., \& {Mo}, H.~J. 2006,
  MNRAS, 366, 2

\bibitem[{{Wilman} {et~al.}(2005{\natexlab{a}}){Wilman}, {Balogh}, {Bower},
  {Mulchaey}, {Oemler}, {Carlberg}, {Eke}, {Lewis}, {Morris}, \&
  {Whitaker}}]{Wilman2}
{Wilman}, D.~J., {Balogh}, M.~L., {Bower}, R.~G., {Mulchaey}, J.~S., {Oemler},
  A., {Carlberg}, R.~G., {Eke}, V.~R., {Lewis}, I., {Morris}, S.~L., \&
  {Whitaker}, R.~J. 2005{\natexlab{a}}, MNRAS, 358, 88

\bibitem[{{Wilman} {et~al.}(2005{\natexlab{b}}){Wilman}, {Balogh}, {Bower},
  {Mulchaey}, {Oemler}, {Carlberg}, {Morris}, \& {Whitaker}}]{Wilman1}
{Wilman}, D.~J., {Balogh}, M.~L., {Bower}, R.~G., {Mulchaey}, J.~S., {Oemler},
  A., {Carlberg}, R.~G., {Morris}, S.~L., \& {Whitaker}, R.~J.
  2005{\natexlab{b}}, MNRAS, 358, 71

\bibitem[{{Wilman} {et~al.}(2008){Wilman}, {Pierini}, {Tyler}, {McGee},
  {Oemler}, {Morris}, {Balogh}, {Bower}, \& {Mulchaey}}]{wilmanMIPS}
{Wilman}, D.~J., {Pierini}, D., {Tyler}, K., {McGee}, S.~L., {Oemler}, Jr, A.,
  {Morris}, S.~L., {Balogh}, M.~L., {Bower}, R.~G., \& {Mulchaey}, J.~S. 2008,
  MNRAS, accepted, (arXiv/0802.2549)

\bibitem[{{Wilman et al.}(2008)}]{wilmanMORPH}
{Wilman et al.} 2008, in preparation

\bibitem[{{Wolf} {et~al.}(2005){Wolf}, {Bell}, {McIntosh}, {Rix}, {Barden},
  {Beckwith}, {Borch}, {Caldwell}, {H{\"a}ussler}, {Heymans}, {Jahnke},
  {Jogee}, {Meisenheimer}, {Peng}, {S{\'a}nchez}, {Somerville}, \&
  {Wisotzki}}]{wolf}
{Wolf}, C., {Bell}, E.~F., {McIntosh}, D.~H., {Rix}, H.-W., {Barden}, M.,
  {Beckwith}, S.~V.~W., {Borch}, A., {Caldwell}, J.~A.~R., {H{\"a}ussler}, B.,
  {Heymans}, C., {Jahnke}, K., {Jogee}, S., {Meisenheimer}, K., {Peng}, C.~Y.,
  {S{\'a}nchez}, S.~F., {Somerville}, R.~S., \& {Wisotzki}, L. 2005, ApJ, 630,
  771

\bibitem[{{Yee} {et~al.}(2000){Yee}, {Morris}, {Lin}, {Carlberg}, {Hall},
  {Sawicki}, {Patton}, {Wirth}, {Ellingson}, \& {Shepherd}}]{CNOC2}
{Yee}, H.~K.~C., {Morris}, S.~L., {Lin}, H., {Carlberg}, R.~G., {Hall}, P.~B.,
  {Sawicki}, M., {Patton}, D.~R., {Wirth}, G.~D., {Ellingson}, E., \&
  {Shepherd}, C.~W. 2000, APJS, 129, 475

\end{thebibliography}

\appendix
\section{Low Redshift Group Finding Algorithm}\label{lowfinder}

The primary goal of our low redshift group
finding algorithm is to reproduce the selection criteria applied to our high
redshift groups. Thus, our algorithm is not the most
efficient method possible, but it does accurately reproduce our selection and the possible
biases within. Further, this method will not result in a complete
sample of groups in the MGC strip.

We first find groups in the SDSS main galaxy sample using the original
method of \citet{CNOC2-groups}. The MGC strip is a narrow region, $\sim$ 35
arcmins across, which makes group finding within the strip itself
difficult. For this reason, we first find groups in the SDSS, in a
region 2 degrees across and centered on the MGC.  We define our
SDSS galaxy sample to be directly
analogous to the CNOC2 sample.  There are two areas of particular
relevance to this work where these
differ: completeness and depth.  Because the SDSS has a much higher
completeness ($\sim$ 90$\%$) than the CNOC2 redshift survey ($\sim$ 48
$\%$), we randomly remove half  the SDSS galaxies.  Further,
we use the same absolute magnitude cut as Carlberg et al., $M_R=-18.5$, with an additional
evolution correction of 1 magnitude per unit redshift. 

\begin{figure}
\leavevmode \epsfysize=8cm \epsfbox{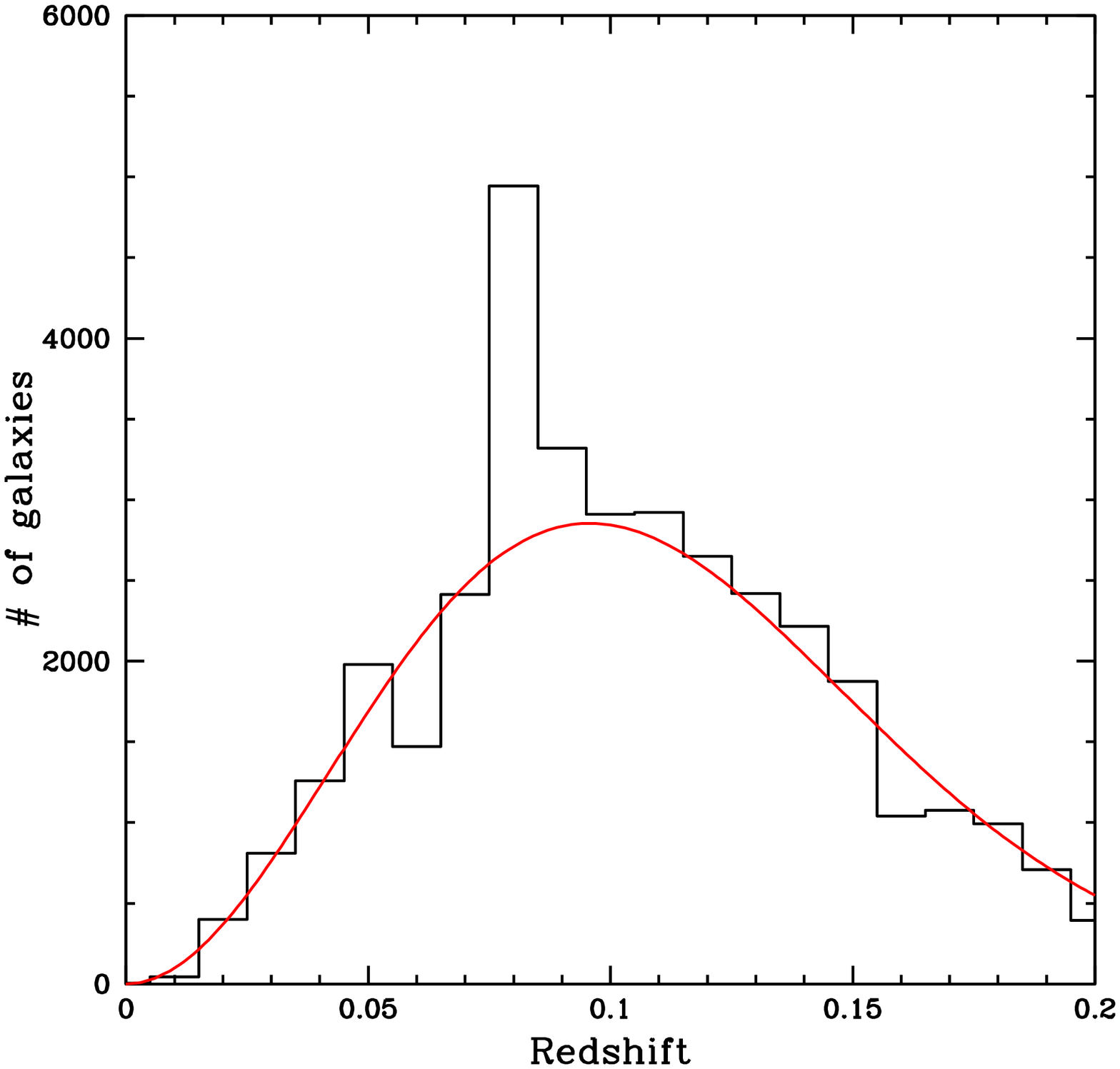}
\caption{The redshift distribution of the full sample of SDSS galaxies
  which were used to find groups at low redshift. The red line is
  the Maxwellian fit to the data which was used to estimate the background.}
\label{ng-sdss}
\end{figure}

Carlberg et al.'s primary goal was to find virialized groups in
overdense environments, so they estimate the overdensity of each
galaxy and restrict their group finding algorithm to galaxies in dense
environments. A cylinder of 0.33\ho\ Mpc radius and $\pm$6.67\ho\ Mpc line-of-sight depth is centered
around each galaxy and the number of galaxies within the cylinder is
counted. If there are fewer than 3 neighbors in this cylinder, the
process is repeated with a cylinder 1.5 times larger. A background estimate is then obtained by randomly drawing
points from a redshift distribution fit to the entire sample. Figure
\ref{ng-sdss} shows the redshift distribution of galaxies in our sample and our analytic
fit. If the number of neighbours in the cylinder is greater than the
background estimate then the main galaxy is kept as a possible group
member. 

Starting with the galaxy with the greatest overdensity, we begin a
trial group by adding any galaxies within the original cylinder, and
any of their friends. When we run out of friends we have a trial
group, for which the geometric position, redshift and velocity dispersion are
computed. Galaxies are trimmed or added within $1.5R_{200}$ and three velocity
dispersions, where $R_{200}$ is the radius at which the density is 200
times the critical density. This process is iterated four times with the
requirement that the last two
iterations are identical. A group is moved to the next stage if it has
more than two members. This concludes the Carlberg et al. algorithm. 

The next stage is to emulate the process in \citet{Wilman1}, to
account for the targeted spectroscopic follow-up, which resulted in a
more nearly complete redshift sampling around selected groups. We do this by including
the complete SDSS catalogue. We use the Carlberg initial centres but
set the velocity dispersion equal to 500 km/s, as was done by Wilman
et al. This was done to remove any bias in the starting velocity
dispersions, which were only based on very few galaxies. We again iterate on
these positions using the entire SDSS catalogue and recompute the
luminosity weighted centres. We compute the velocity dispersion at
each step using the Gapper estimator and remove galaxies outside
two velocity dispersions and 500\ho kpc. Finally, we keep only those groups
which lie completely within the MGC strip. Using this method we have
19 groups with velocity dispersions between 100 km/s and 700
km/s, and which lie within 0.04 $<$ z $<$ 0.12. 
 
There exist a large number of group and cluster catalogues based
on the SDSS and 2DF surveys with which we can compare. This is especially important to calibrate
the systematic effects which may be present in our high redshift
sample, which doesn't have sufficient completeness to quantify within the survey itself. One of the more
popular group finding algorithms, and the most direct analogue to our
method, is that of \citet{Berlind}.  They use a traditional
friends-of-friends algorithm in position-redshift space to find groups
in three different 
volume limited samples. We find that 12 out of 13 of our groups below
z=0.1, the depth of the deepest Berlind sample, are also found in the
Berlind catalogue, ie. the Berlind group centres are contained within
our groups. Eighty-five of the 99 group members which make up the low
redshift group sample in this paper
would also be group members if we were to use the Berlind catalogue as
our group catalogue. 

\section{Dependence of galaxy properties on PSF size} \label{mgc_red_depend}

As discussed in \textsection \ref{comparesur}, the CNOC2 and MGC
surveys compare well in absolute surface brightness limits, physical
size of the PSF and the rest waveband used in the morphological
decomposition.  In fact, the biggest variation in these parameters is
actually within the MGC sample itself. The PSF size is $\sim$ 3 kpc in size
at z = 0.14 and just $\sim$ 1 kpc at z=0.05.

In this section, we investigate the redshift dependence of the key
morphological indicators in three bins of redshift within the MGC
sample. Some care must be taken because, as we have seen, the disk
fraction changes rapidly with luminosity; therefore, without first matching on
luminosity we would have a higher disk fraction in the lowest redshift
bins. We have broken the sample into  low (0 $<$ z $<=$ 0.05), medium
(0.05 $<$ z $<=$ 0.1) and high (0.1 $<$ z $<$ 0.15) redshift bins. Each
galaxy in the 0.05 $<$ z $<=$ 0.1 bin was randomly matched to a galaxy in
each of the other two redshift ranges with an absolute B
magnitude within 0.03 magnitudes. The sample has 2169 galaxies in the $0.05<z<0.1$
range. They were matched to a unique sample of 266 galaxies in the $0<z<0.05$ range and 1354 galaxies in the $0.1<z<0.15$
range. Each galaxy was weighted by the number of times it was matched. 

\begin{figure}
\leavevmode \epsfysize=8cm \epsfbox{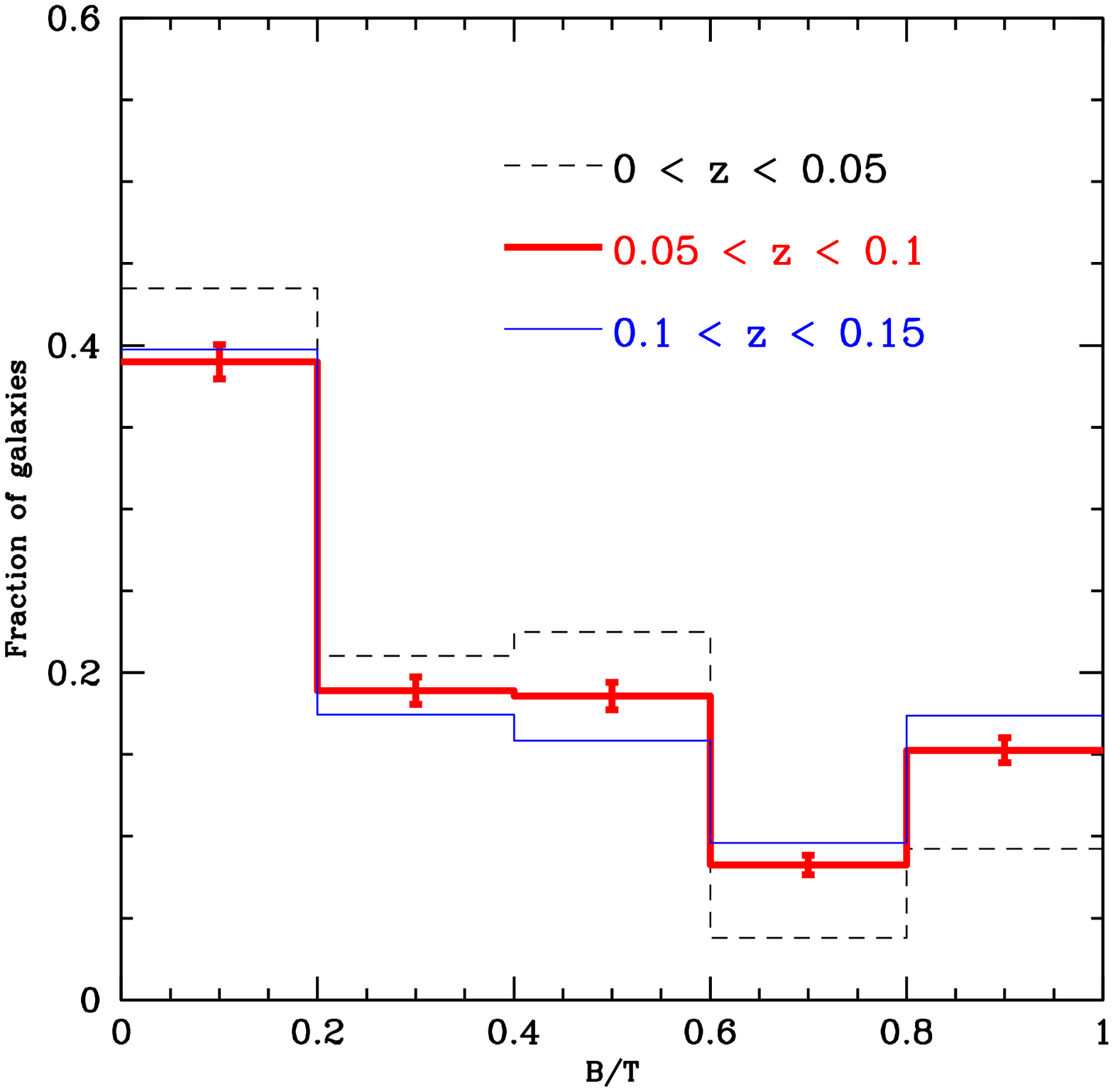}
\caption{The B/T distribution of luminosity-matched samples of MGC
  galaxies in bins of redshift. The sample is divided into  low (0 $<$
  z $<$ 0.05; thin, dashed, black line), medium (0.05 $<$ z $<$ 0.1;
  thick red line) and
  high (0.1 $<$ z $<$ 0.15; thin, solid, blue line) redshift bins}
\label{bt_redbin}
\end{figure}

Figure \ref{bt_redbin} shows the B/T distribution of the three different
redshift samples. These distributions are very similar in all redshift
bins and our conclusions are unchanged if the field sample is taken as
any of these bins. 

\begin{figure}
\leavevmode \epsfysize=8cm \epsfbox{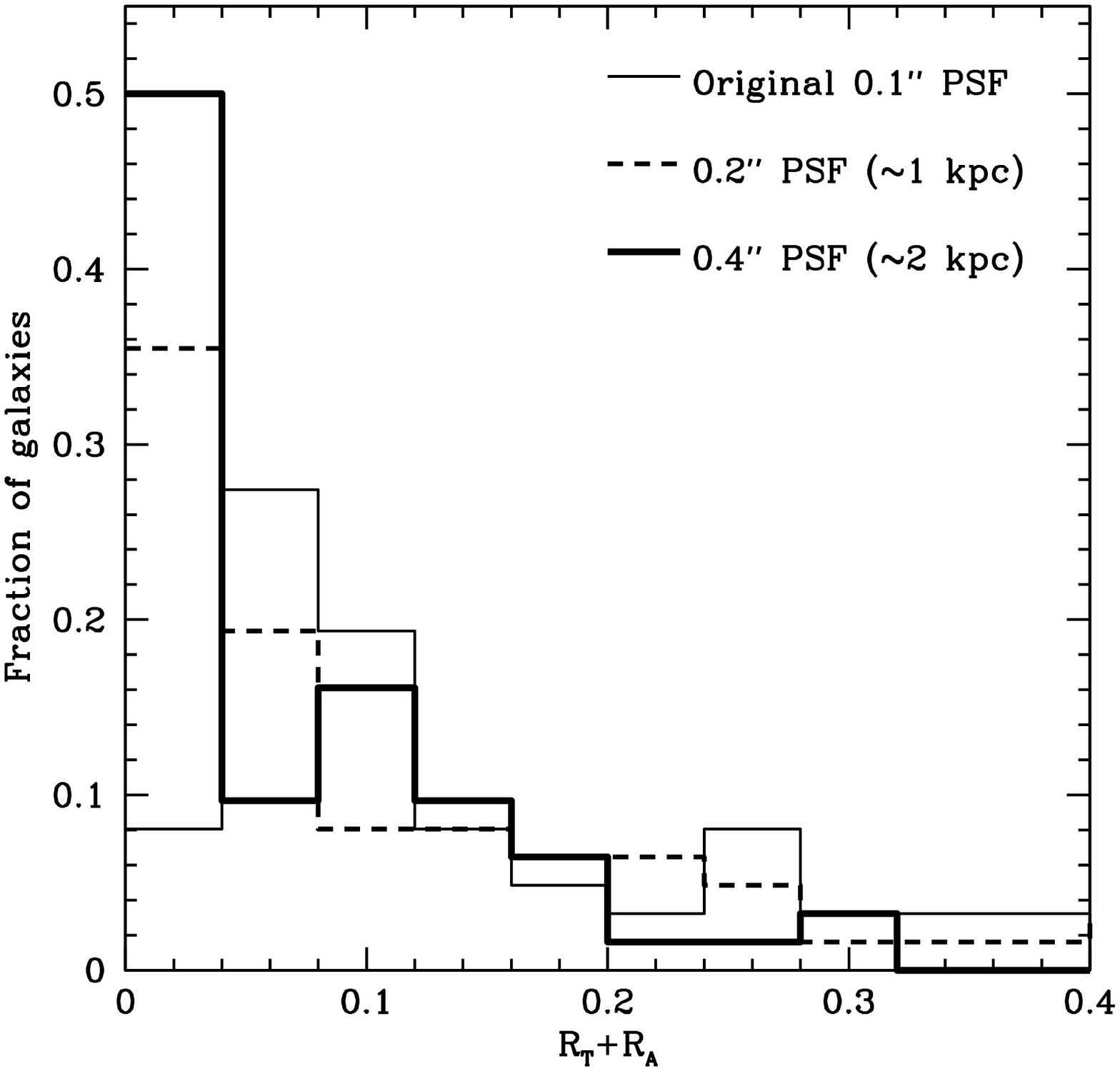}
\caption{The distribution of asymmetries (R$_T$+R$_A$) measured within 2
  halflight radii for different physical resolutions for a sample of
  60 representative CNOC2 galaxies. The images were
  blurred with a Gaussian with a PSF of 1 (black, dashed line) and 2
  kpc (thick, solid black line). The original, unblurred asymmetry is
  shown as the thin black line.}
\label{asym_blur}
\end{figure}

In \textsection \ref{asym}, we claim that the observed
 difference in the mean asymmetries between the CNOC2 sample and the
 MGC sample is due to the better physical resolution of the CNOC2 images. 
To test this we reduce the resolution of the original image and the
 residual image of a representative sample of 60 galaxies from our
 CNOC2 sample by convolving with a Gaussian with different widths. Figure
\ref{asym_blur} shows the asymmetry distribution of the original
 sample (thin, solid black line), and the asymmetries after broadening
 with a 1 kpc (black, dashed line) and 2 kpc PSF (thick, solid black
 line). Clearly the asymmetry is reduced with poorer physical
 resolution, which explains the higher asymmetries of the CNOC2
 sample. For this reason, we do not compare the asymmetries between the
 two surveys and always use samples matched in redshift within a given
 survey.

\section{Images of a Sample of Group and Field galaxies at z=0.4} \label{thumbs}
\begin{figure*}
\leavevmode  \epsfbox{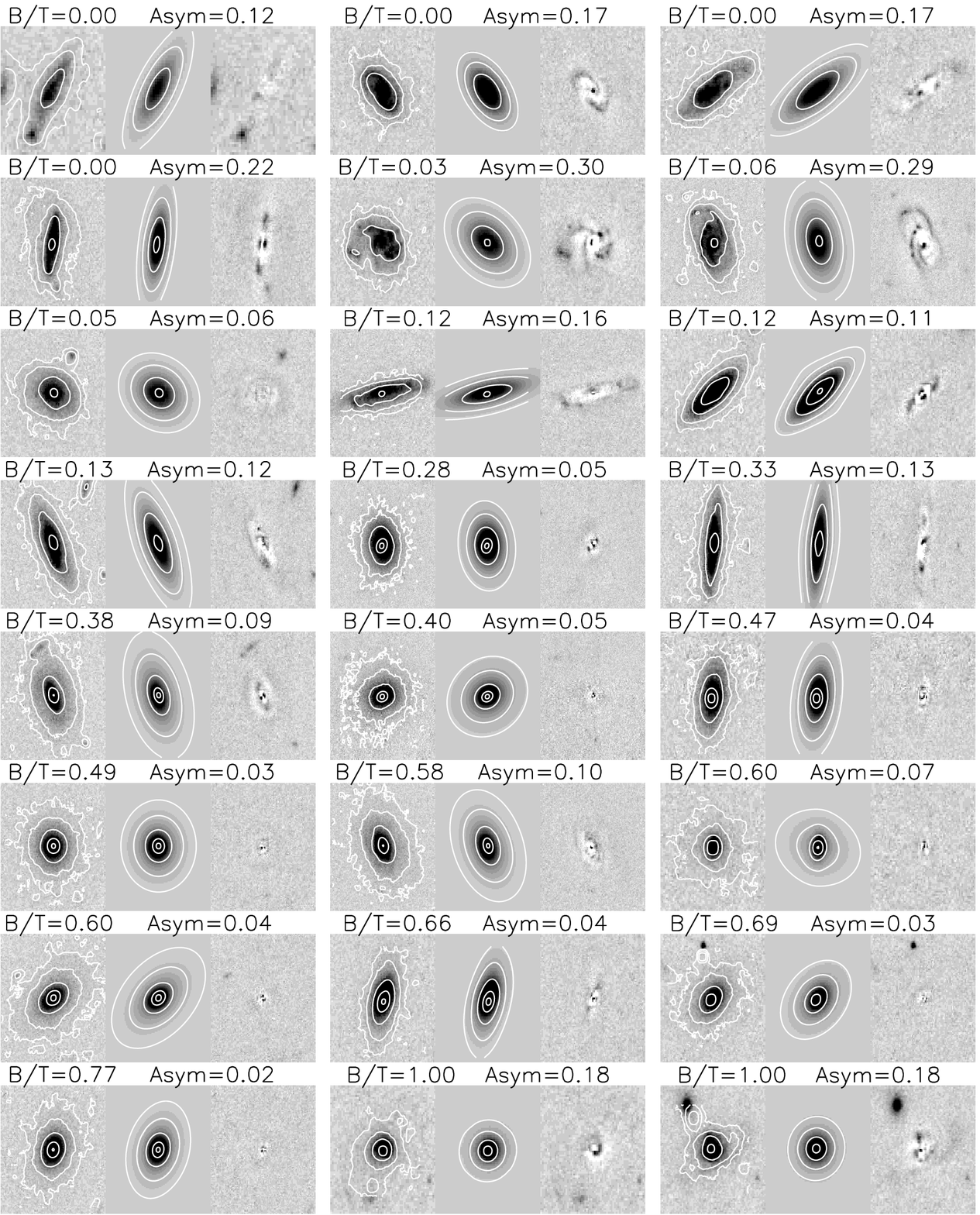}
\caption{A representative sample of group galaxy images from the z=0.4
  sample. For each of 24 galaxies is the {\it HST} ACS image (left panel),
  the \textsc{gim2d} output model galaxy (middle panel), and the
  residual of the {\it HST} ACS image after the model is subtracted (right
  panel). Each galaxy is listed with the Bulge to total ratio (B/T) and Asymmetry
  (Asym) computed by \textsc{gim2d}. }
\label{group_thumb}
\end{figure*}

\begin{figure*}
\leavevmode  \epsfbox{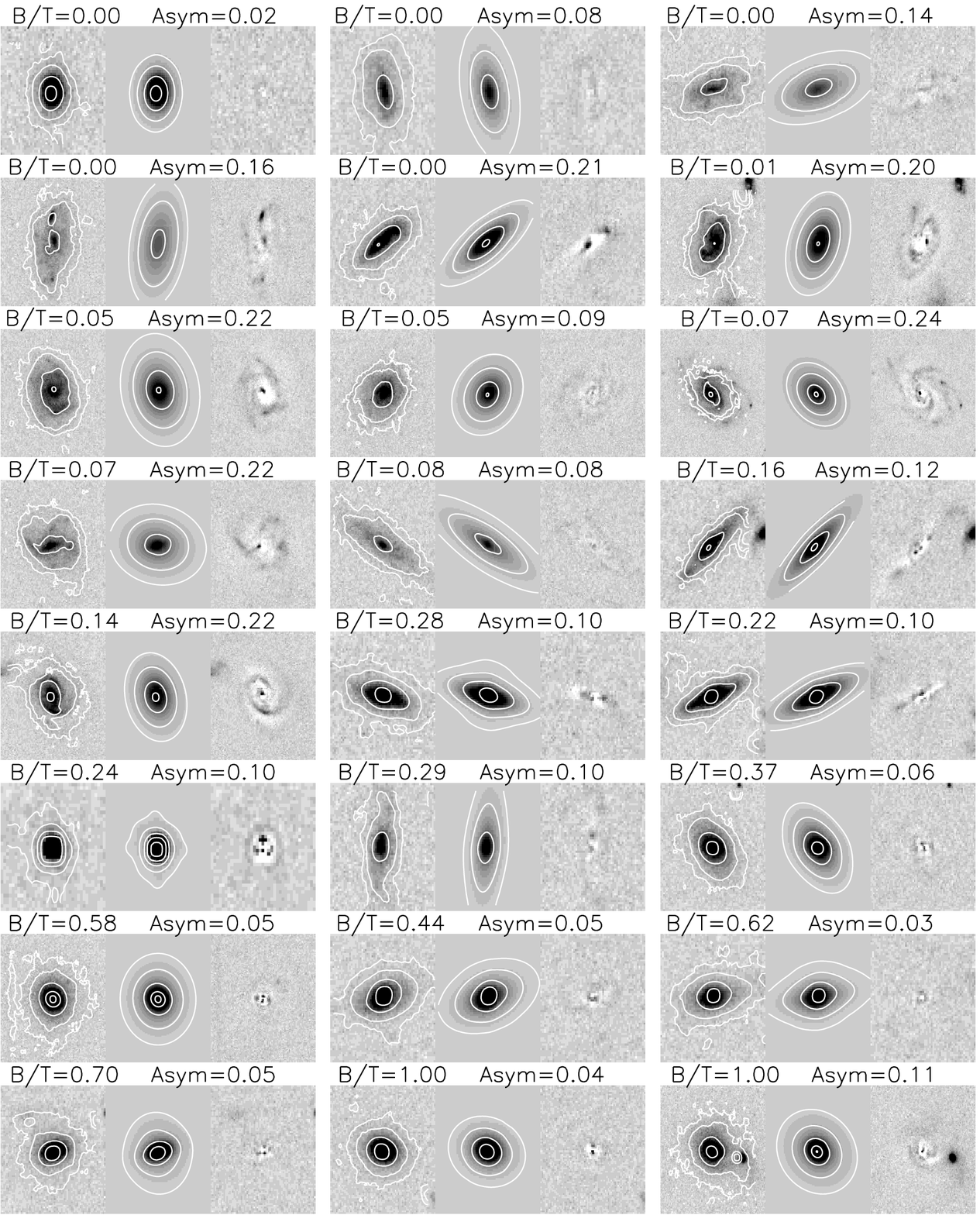}
\caption{As in Figure \ref{group_thumb}, but for a representative
  sample of 24 field galaxies in the z=0.4 sample.}
\label{field_thumb}
\end{figure*}

\begin{figure*}
\leavevmode  \epsfbox{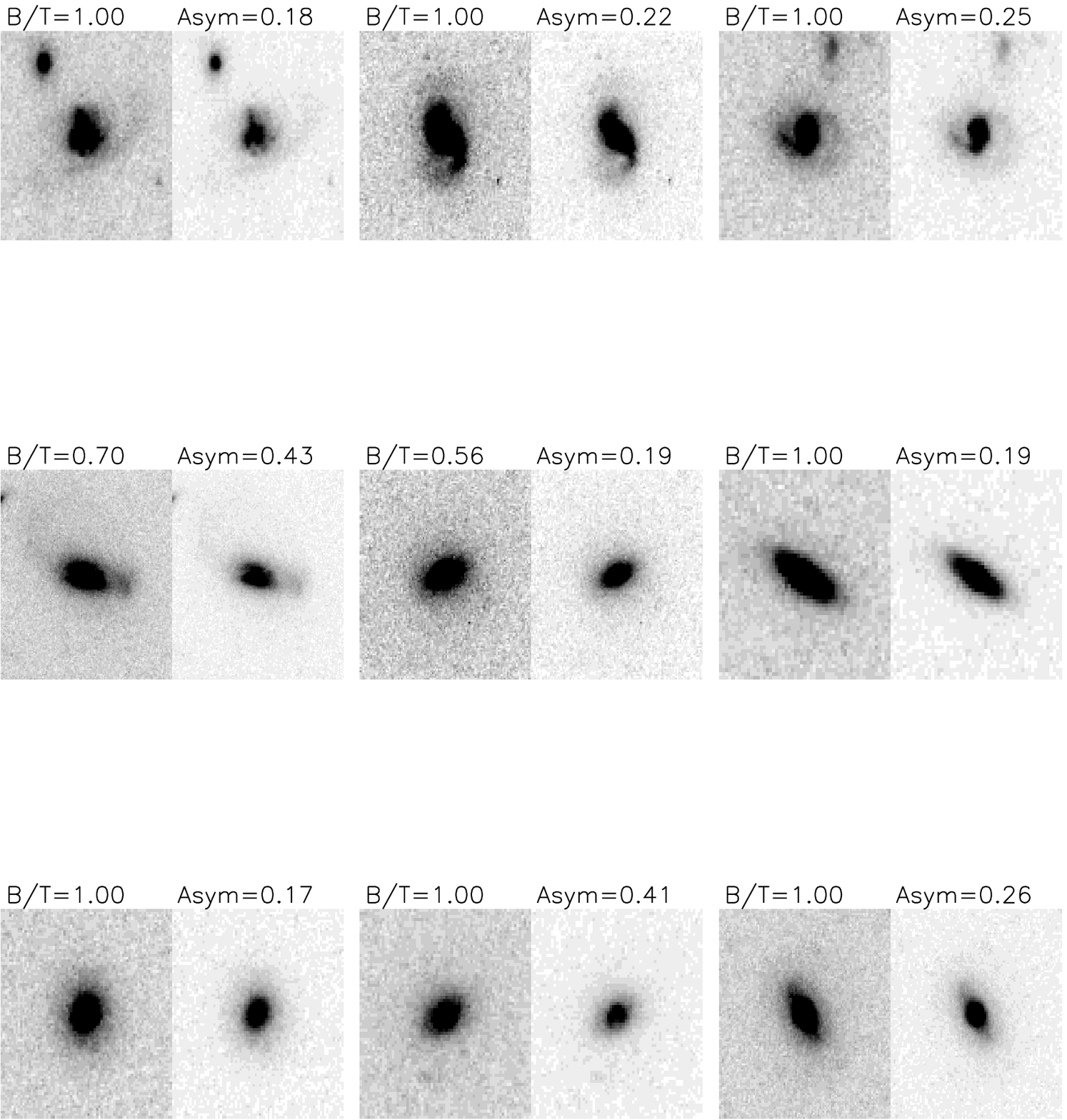}
\caption{The nine group galaxies in the z=0.4 sample which are bulge
  dominated (B/T$>$0.5) and highly asymmetric ($R_T~+~R_A >$
  0.16). The galaxies are shown with two stretches to show the
  interaction features. Prominent interaction features are seen in the
  first four of these galaxies.}
\label{hiasym_thumb}
\end{figure*}

In this section, we show a representative sample of the group and
field galaxies from our CNOC2 z=0.4 sample. As discussed in
\textsection \ref{comparesur}, the thumbnail images are from {\it HST} ACS 
observations. Each image is shown together with the \textsc{gim2d}
model and the residual of the {\it HST} image after the model was
removed. We show the group galaxies in Figure \ref{group_thumb}, and
the field galaxies in
Figure \ref{field_thumb}. In Figure \ref{hiasym_thumb}, we show images
of the nine group galaxies in the CNOC2 sample which are bulge
dominated (B/T$>$0.5) and have high asymmetries  ($R_T~+~R_A
>$0.16). The first four of these galaxies show interaction features.

\end{document}